\documentclass[twocolumn]{aa} 
\usepackage{orcidlink}

\usepackage{graphicx}
\usepackage{txfonts}
\begin{document}

   \title{Long-period maser-bearing Miras in the Galactic center}

   \subtitle{Period-luminosity relations and extinction estimates
   \thanks{Tables 1 and 2 are available in electronic form at the CDS via anonymous ftp to cdsarc.cds.unistra.fr (130.79.128.5) or via \url{https://cdsarc.cds.unistra.fr/cgi-bin/qcat?J/A+A/}
   }
   }

   \author{M. O. Lewis 
          \orcidlink{0000-0002-8069-8060} 
          \inst{1}
          \and
          R. Bhattacharya 
          \orcidlink{0009-0007-8229-3036}
          \inst{2}
          \and
          L.O. Sjouwerman 
          \orcidlink{0000-0003-3096-3062}
          \inst{3}
          \and
          Y. M. Pihlstr\"om 
          \orcidlink{0000-0003-0615-1785}
          \inst{2}\fnmsep\thanks{YMP is also an Adjunct Astronomer at the National Radio Astronomy Observatory}
          \and
          G. Pietrzy\'nski
          \orcidlink{0000-0002-9443-4138}
          \inst{1}
          \and
          R. Sahai 
          \orcidlink{0000-0002-6858-5063}
          \inst{4}
          \and
          P. Karczmarek
          \orcidlink{0000-0002-0136-0046}
          \inst{5,1}
          \and
          M. G\'orski
          \orcidlink{0000-0002-3125-9088}
          \inst{1}
          }

   \institute{Nicolaus Copernicus Astronomical                Center, Polish Academy of Sciences
              Bartycka 18, 00-716 Warszawa, Poland
         \and
             Physics and Astronomy Department, University of New Mexico
             87131 Albuquerque, NM, USA
        \and
            National Radio Astronomy Observatory
             Pete V.\ Domenici Science Operations Center Operations Center
            Socorro, NM 87801, USA
        \and
             Jet Propulsion Laboratory, MS 183-900, California Institute of Technology, Pasadena, CA 91109, USA
        \and 
            Departamento de Astronom{\'i}a, Universidad de Concepci{\'o}n, Casilla 160-C, Concepci{\'o}n, Chile
            }

   \date{Received April 03 2023; accepted July 25 2023}

 
  \abstract
   {We establish a sample of 370 Mira variables that are likely near the Galactic center (GC). The sources have been selected from the OGLE and BAaDE surveys based on their sky coordinates, OGLE classifications, and BAaDE maser-derived line-of-sight velocities.
   As the distance to the GC is known to a high accuracy, this sample is a test bed for reddening and extinction studies toward the GC and in Mira envelopes. 
   We calculated separate interstellar- and circumstellar-extinction values for individual sources, showing that there is a wide range of circumstellar extinction values (up to four magnitudes in the K$_s$ band) in the sample, and that circumstellar reddening is statistically different from interstellar reddening laws. 
   Further, the reddening laws in the circumstellar environments of our sample and the circumstellar environments of Large Magellanic Cloud (LMC) Miras are strikingly similar despite the different metallicities of the samples. 
   Period-magnitude relations for the mid-infrared (MIR) WISE and MSX bands are also explored, and in the WISE bands we compare these to period-magnitude relationships derived from Miras in the LMC as it is important to compare these LMC relations to those in a higher metallicity environment. Emission from the envelope itself may contaminate MIR magnitudes altering the relations, especially for sources with thick envelopes.}


   \maketitle
%

\section{Introduction} \label{sec:intro}
The asymptotic giant branch (AGB) is a late stage of evolution for stars with progenitor masses between about 0.8 and 8 M$_\odot$. As this stage precedes the planetary nebulae phase, it is characterized by large amounts of mass loss ($\sim$10$^{-8}$-10$^{-4}$ M$_\odot$ yr$^{-1}$; \citealt{hofnerandolofsson}), causing these stars to be surrounded by circumstellar envelopes (CSEs). These CSEs have a large effect on the observed optical and near-infrared (NIR) properties of AGB stars, obscuring much of the light at these wavelengths. Additionally, AGB stars are intrinsically variable, typically with long periods (hundreds of days), and based on their optical and NIR light curves can be categorized as semi-regular, irregular, or Mira long-period variables \citep{habing96}. 
\par Mira variable stars in particular have large amplitudes (greater than 0.8 mag in the I band or greater than 0.4 mag in the K band; \citealt{soszynski13,whitelock03}) and oscillate in the fundamental mode \citep{wood96}; they have been shown to follow color-period and period-magnitude relations in the NIR in a number of studies and environments \citep{whitelock00, whitelock03, rejkuba04, whitelock08, matsunaga12, yuan17b, bhardwaj19, lebzelter22}.
However, these relationships are often difficult to interpret when extinction (either interstellar or circumstellar) is high, and often also become more difficult to define for samples with periods longer than $\sim$450 days \citep{ita11, qin18, matsunaga09, yuan17b}. While some of the uncertainty associated with the period-luminosity relation (PLR) for longer-period Miras can be addressed by using quadratic or kinked (with a break in the slope) relations, which account for hot-bottom burning \citep{whitelock03}, the difficulty in constraining Mira PLRs at long periods is also related to CSE extinction issues. In general, longer period Miras correspond to redder stellar colors and thicker CSEs \citep{vdvH88}, and thus the long-period end of such relations is most affected by the presence of a thick CSE. Despite these limitations, Mira PLRs in the Large Magellanic Cloud (LMC) have shown low scatter (e.g., 0.12 mag; \citealt{yuan17b}), demonstrating that these relations are suitable for distance determination. Although establishing a sample of Mira variables in the Milky Way (MW) with known distances is more difficult than using LMC samples, Mira PLRs in the MW appear, so far, to be similar to those in the LMC \citep{glass95, whitelock00, feast02}. More recently, \cite{goldman19} found no dependence on metallicity for AGB pulsations in very metal-poor galaxies, but a slightly steeper PLR has been found in the MW as compared to LMC relations using Gaia parallaxes \citep{sanders23}. MW Mira PLRs remain of great interest, as solidifying the effect (or lack of effect) of metallicity on these relations is an important step for bolstering their use as distance estimators. In particular, in this study we are interested in Miras in the high-metallicity, bulge region of the MW. 
\par While PLRs (or period-magnitude relations) are best known as tools for calculating distances, these relations can also be used to calculate reddening and extinction in samples at a given distance, for example, the Galactic center (GC; e.g., \citealt{ita11, qin18}).
Samples of Mira variables have been used to calculate both the distance to and the reddening and extinction associated with the GC \citep{schultheis09, matsunaga09, ita11, qin18}. 
These samples often include hundreds of Miras and are based on assuming: i) that the center of the Mira population is colocated with the GC and ii) that the Mira sample is symmetrically distributed on either side of the GC. 
Similar studies have been done with Cepheid variables \citep{bhardwaj17, dekany19}. 
\par To form cleaner samples of GC stars, kinematic criteria can be used, effectively removing all sources that may be associated with the disk.
We note that such criteria may also remove a fraction of bulge sources, but with the net result of a much more homogeneous sample.
Line-of-sight velocity measurements via optical and NIR spectroscopy are very difficult to obtain close to the plane and in the GC; however, radio-wavelength observations of masers are unaffected by extinction and available for thousands of AGB stars in the plane (e.g., \citealt{blommaert94, messineo02, deguchi04}). 
Specifically, SiO maser emission in stellar envelopes allows for measurements of line-of-sight velocity of sources regardless of obscuration by gas and dust (e.g., \citealt{deguchi04, messineo18}). 
As such, many samples of Mira stars with available line-of-sight velocities are comprised primarily of maser-bearing Mira stars, which do not necessarily reflect the properties of general MW Mira variables. 
Nevertheless, maser-bearing Mira variables play an important role in establishing MW PLRs because the distances to these sources can potentially be measured via maser parallaxes. Therefore, it is critical to understand how maser-bearing Miras differ from the general population in terms of CSE extinction and the corresponding effects on PLRs.  
Herein we address a sample of maser-bearing Mira variables near the GC.

\par Section 2 describes the sample selection and the surveys from which data are collected. Section 3 presents our results and starts by showing that our sample consists of especially long-period Miras.
This section also includes the derivation of interstellar and circumstellar extinctions toward GC sources and presents period-magnitude diagrams for wavelengths 0.9 - 21 $\mu$m. 
In Section 4 we discuss mid-infrared (MIR) period magnitude relations in the sample showing that these stars are brighter at a given period than their LMC counterparts.
We also explore the correlations between circumstellar extinction in the NIR bands showing these correlations are not what one would expect if circumstellar and interstellar reddening laws were the same. Finally in this section, we show several correlations between K-band extinction and various IR colors. Conclusions and a summary are presented in Section 5.

\section{Data and sample selection} \label{sec:sample}

\par 
The Mira variable sources discussed in this work are adopted from the OGLE Mira catalog \citep{iwanek22} and the BAaDE SiO maser catalog (\citealt{lewis2021}, Sjouwerman et al.\,in preparation) as described below. 
OGLE Bulge variables are identified from within the OGLE Bulge field and include 40,356 Mira variables (65,981 Miras in the Bulge and Disk OGLE data combined; \citealt{iwanek22}). Photometry for this catalog consists mostly of Cousins I-band data with some V-band points. The median number of epochs per source is over 200 in the I band, and some sources have been monitored for as long as 20 years with over 16,000 epochs of I-band data.
The identification of long-period variables, and specifically Miras, within the OGLE survey is described in \cite{iwanek22} and references therein. The methodology relied largely on visual inspections of periodograms and light curve shapes. Periods between 10 and 2000 days are considered, and the final period is confirmed via multiple methods \citep{iwanek22}. No formal errors on these periods are given. Within the long-period variables, Miras are classified by the amplitudes of their light curves---which are greater than 0.8 magnitudes in the I band.  
\par 
The BAaDE survey is a radio-wavelength spectral survey of 43 GHz J=1--0 and 86 GHz J=2--1 SiO maser transitions in AGB stars. 
Roughly 20,000 IR-color-selected stars have been observed by the VLA and ALMA, and SiO maser emission has been detected in 56\% of these sources (we note that this rate is artificially lowered by contamination from young stellar objects and carbon-rich AGB sources; \citealt{lewis20a}), yielding maser flux and line-of-sight velocity information for these sources.
Another 8,000 ALMA sources will be analyzed in the coming year. BAaDE line-of-sight velocities are estimated to be accurate to $\sim$2 km s$^{-1}$ and are available for all sources where an SiO maser is detected. The SiO-detection criterion also ensures that the sample with reliable velocities is composed (likely exclusively) of oxygen-rich (O-rich) stars. Relying on maser data for these line-of-sight velocities is necessary in large parts of the Bulge and the plane where even K-band extinction can range from 1 to 3.5 mag \citep{gonzalez12}, greatly affecting the availability of optical and IR spectra; this is especially true in the GC sample presented here.
\par The OGLE Bulge Mira catalog was cross-matched with the BAaDE catalog, in order to create a sample of sources with known variability type, periods, and line-of-sight velocities. 
The cross-match was performed using only sky-coordinate information with a 3\arcsec\ matching radius. The coordinates used for the BAaDE sources are those from the 2MASS catalog \citep{skrutskie06}, where available, or the MSX catalog \citep{egan03} otherwise.
This is because 2MASS positions more accurately align with the maser positions than the MSX positions \citep{ylva}.
The cross-match returned 2,902 Mira variables in both the OGLE and BAaDE catalogs. 
The Vista Variables in the V\'ia L\'actea data was not used to obtain variability information as many Miras are saturated in that catalog \citep{nikzat22}.
Of Mira variables in both the OGLE and BAaDE catalogs, 2,437 have line-of-sight velocities available from BAaDE data, while the rest did not show SiO maser activity during the BAaDE observations. 
The BAaDE maser detection rate among OGLE Mira variables is therefore 84\%. This is a notably high fraction given that SiO maser sources are variable and that both \cite{michael18} and \cite{ylva} detect SiO masers in roughly 80\% of sources when reobserving samples of known SiO maser stars.
\par The sample has also been matched to the 2MASS, WISE, and MSX catalogs (the BAaDE catalog is derived from 2MASS and MSX, and the WISE match was conducted separately) and we therefore have access to 2MASS J (1.2 $\mu$m), H (1.6$\mu$m), K$_s$ (2.2$\mu$m); WISE 1 (3.4$\mu$m), 2 (4.6$\mu$m), 3 (12$\mu$m); and MSX A (8$\mu$m), C (12$\mu$m), D (14$\mu$m), and E (21$\mu$m) magnitude measurements \citep{skrutskie06, wright10, egan03}. Only 2MASS matches with quality flags A or B were included, and sources flagged as extended or artifacts were further excluded. WISE matches with quality flags U and X were excluded. Given the high quality of the IR data once these flags are considered and the brightness of AGB stars, we find that blending is unlikely to have much affect on this data even in crowded regions. 2MASS measurements are single-epoch observations of variable sources, as opposed to average magnitudes over the stellar phase, which will contribute to scatter and uncertainty when comparing to period-IR relations. WISE measurements are averaged over several epochs and may suffer less from this effect, but still do not represent average magnitudes derived from light-curve fitting.
\par Finally, we define a sample of OGLE/BAaDE sources with absolute line-of-sight velocities over 100 km s$^{-1}$, absolute Galactic longitudes less than 3$^{\circ}$, and absolute latitudes below 4$^{\circ}$. Six sources (ad3a-05111, ad3a-03680, ad3a-04158, ad3a-03047, ad3a-01580, and ad3a-04121) which fit these criteria are not included in the sample because their light curve coverage is sparse (especially as compared to other OGLE Mira variables), making period determinations unreliable. 
This resulting sample of 370 sources is presented as our GC sample.
This sample is assumed to be in the inner bulge, near the center of MW, that is, near Galactic latitude and longitude 0$^\circ$ and near a distance of 8 kpc \citep{reid19, gravity19}. 
By using this relatively small, carefully selected sample, we avoid the need to make assumptions about the symmetry of the Milky Way Mira population about the GC or the coincidence of the center of the Galactic Mira population and the GC, as was necessary in previous studies regarding the GC distance \citep{schultheis09, matsunaga09, matsunaga12, qin18}. 
Instead we use the new radio-wavelength measurements of the line-of-sight velocities of Miras from BAaDE to select only Miras with high absolute velocities. 
By adding velocity information we exclude sources that are likely to be in the disk, as such sources are expected to have much lower line-of-sight velocities due to their lower velocity dispersion \citep{robin17, ness16}. 
Cutting low-absolute-velocity sources will certainly also exclude many bulge sources, because the average velocity in both the bulge and disk toward the GC is close to zero \citep{mcginn89, trippe08, kunder12}.
However, there is also a large population of bulge sources with high absolute line-of-sight velocities (up to $\sim$300 km s $^{-1}$) which occupy the edges of the velocity distribution \citep{mcginn89, lindqvist92, messineo02, trippe08, kunder12}. These high-velocity sources are expected only in the bulge because the velocity dispersion is much higher (130 km s$^{-1}$; \citealt{valenti18}) than in the disk (40 km s$^{-1}$; \citealt{valenti18}). 
Thus, after our velocity cut, few to no disk sources are expected to remain in the sample, leaving almost exclusively bulge sources. 
We choose to use this smaller sample, which doubtlessly has excluded many bulge stars, as it should have very little contamination from stellar disk populations.
We acknowledge that halo stars may also have higher velocities similar to bulge stars, but we note that the density of halo stars to bulge stars is very low in the plane. That is, we cannot exclude that individual halo stars may be included, but even in the event that such low-metallicity stars manage to sustain SiO masers, the number of these stars in our sample would be extremely low.

\begin{figure*}
    \centering
    \includegraphics[width=0.7\textwidth]{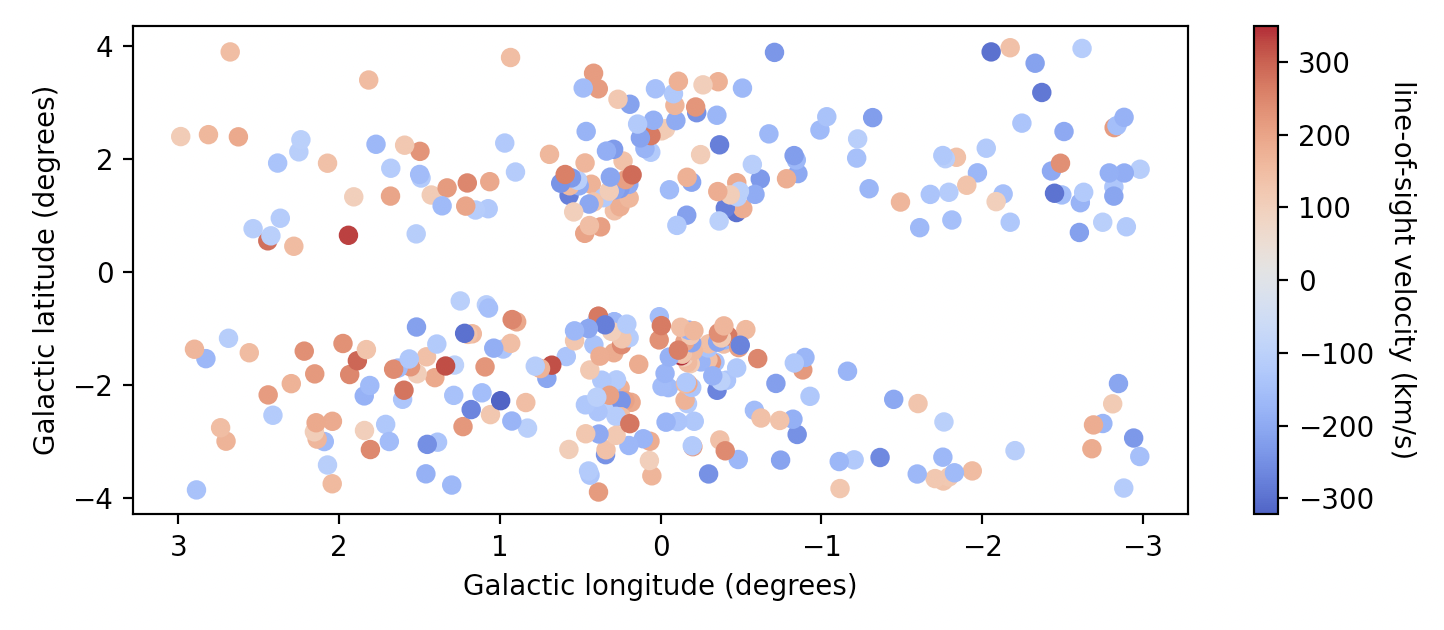}
    \caption{Positions and line-of-sight velocities of the 370 ``Galactic Center" Mira sources, which we assume to be at a distance of 8 kpc. The latitude gap traces areas of high extinction, where the OGLE survey is unable to obtain light curves---although maser information is available in this region.}
    \label{fig:gcsours}
\end{figure*}

Of course this sample selection method depends on a different set of assumptions, namely that high velocity ($>$100 km s$^{-1}$) Mira variables within 4 degrees of l,b=0,0 are at the distance of the GC, i.e., 8 kpc from the Sun (i.e., \citealt{gravity19, reid19, do19, iwanek23}). 
We note that reliable Gaia data is sparse for this sample given the proximity to the Galactic plane and the large distances. Further, Gaia parallaxes do not lead straightforwardly to accurate distances for Mira variables \citep{andriantsaralaza22}; thus Gaia parallaxes are not employed. We also did not adopt the period-based distance estimates for OGLE Miras given by \cite{iwanek23} as these are unlikely to be accurate for highly obscured stars. 
See Sect.\,\ref{sect:error} for discussion of how this distance assumption affects our uncertainty.
The source names from both catalogs, 2MASS positions (i.e., those adopted by the BAaDE survey), maser-derived line-of-sight velocities, and I-band periods are given in Table \ref{tab:basics}. Further, the positions and line-of-sight velocities of the GC sources are shown in Fig.\ref{fig:gcsours}.

\begin{table*}
    \caption{GC sample properties.}
    \centering
    \begin{tabular}{c|c|c|c|r|c}
     BAaDE name & OGLE name & RA & Dec & V$_{\rm{LOS}}$ & period \\
      &  & (J2000) & (J2000) & (km s$^{-1}$) & (days) \\
ad3a-04238 & OGLE-BLG-LPV-251997 & 17:50:27.528 & $-$28:18:39.96 & $-$107 & 504.0 \\
ad3a-04347 & OGLE-BLG-LPV-248279 & 17:43:25.800 & $-$28:12:12.96 & 210 & 405.5 \\
ad3a-04350 & OGLE-BLG-LPV-138399 & 17:57:14.688 & $-$28:12:09.72 & 248 & 455.6 \\
ad3a-04376 & OGLE-BLG-LPV-248715 & 17:44:06.456 & $-$28:10:45.12 & 201 & 362.7 \\
ad3a-04404 & OGLE-BLG-LPV-177264 & 18:01:09.408 & $-$28:08:48.12 & $-$118 & 456.4 \\
ad3a-04412 & OGLE-BLG-LPV-253293 & 17:53:00.000 & $-$28:08:15.00 & $-$218 & 482.2 \\
ad3a-04415 & OGLE-BLG-LPV-252053 & 17:50:34.248 & $-$28:08:11.04 & $-$105 & 524.3 \\
ad3a-04418 & OGLE-BLG-LPV-248322 & 17:43:30.312 & $-$28:07:55.56 & 147 & 273.0 \\
ad3a-04425 & OGLE-BLG-LPV-247533 & 17:42:07.776 & $-$28:07:36.12 & 140 & 405.1 \\
ad3a-04433 & OGLE-BLG-LPV-254833 & 17:56:10.656 & $-$28:07:17.04 & 293 & 557.0 \\
    \end{tabular}
    \tablefoot{BAaDE and OGLE names for the 370 GC sample sources, as well as source coordinates, maser-derived line-of-sight velocities, and I-band periods. Listed positions are those from 2MASS, accurate to within 0.015$\arcsec$ \citep{skrutskie06} and generally within 1$\arcsec$\ of the maser emission \citep{ylva}, which are the coordinates used during the BAaDE survey. Coordinates of the corresponding OGLE source may be different by as much as 3$\arcsec$. Uncertainties on the velocities are $\pm$2 km s$^{-1}$ for all sources and are dominated by the spectral resolution of the BAaDE survey. The full table is available in the online journal. The first ten rows are presented here to demonstrate the form and content. }
    \label{tab:basics}
\end{table*}

\section{Results}
We present results concerning the BAaDE/OGLE cross-match showing that maser-bearing sources have distinctly long periods, and further focus on the 370 GC sources which we assume to be at a distance of 8 kpc based on their sky positions and velocities. These sources allow us to estimate circumstellar extinction on a source-by-source basis by comparing to NIR Mira period-magnitude relations, which reveals that circumstellar extinction can account for several magnitudes of extinction in the K$_s$ band. 
Finally, because extinction is still prominent in the NIR, we move to MIR period-magnitude relations, which are apparent in our sample when interstellar (but not necessarily circumstellar) extinction is corrected for. 

\subsection{Periods}

\begin{figure}[t]
    \centering
    \includegraphics[width=.48\textwidth]{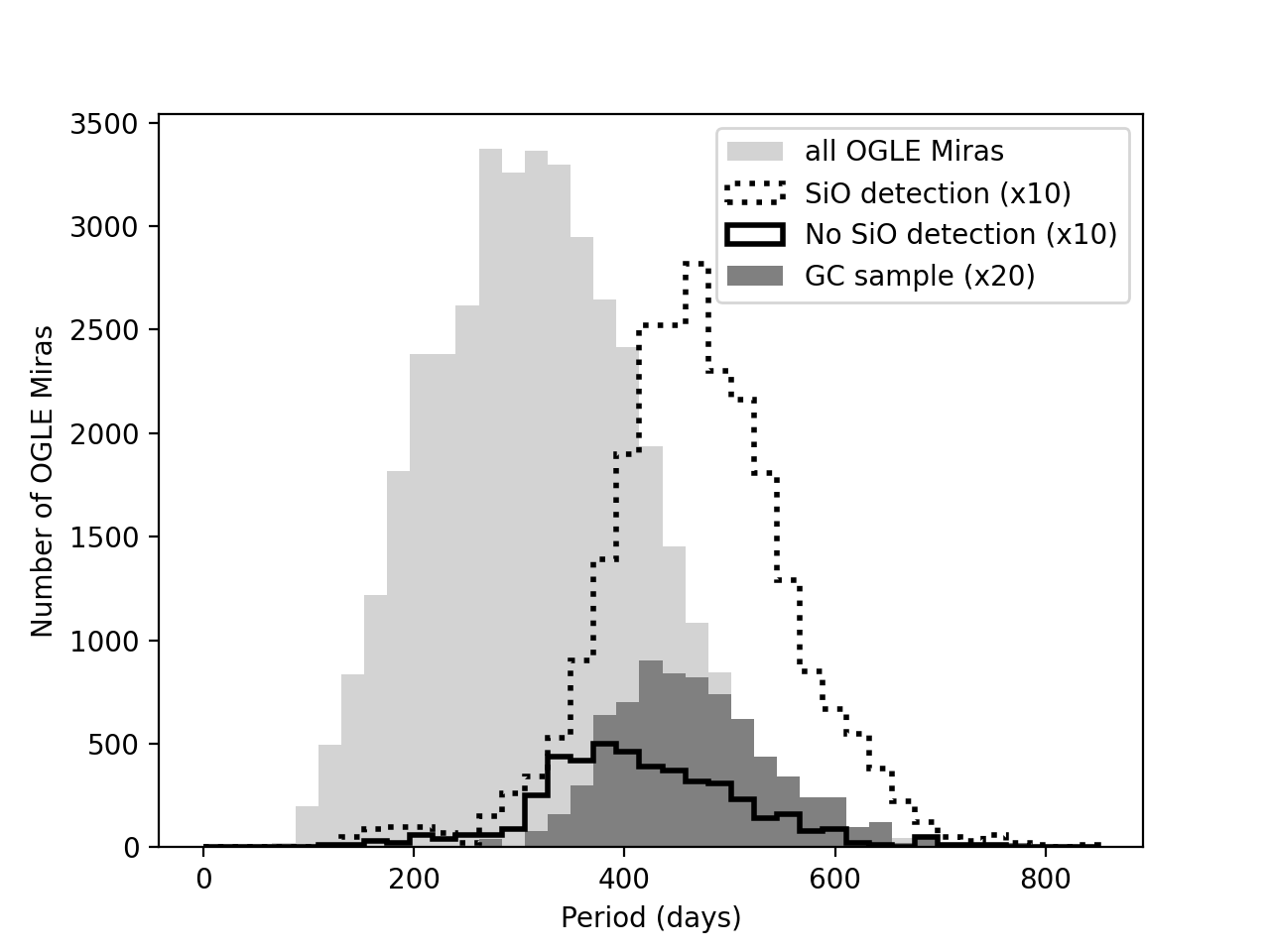}
     \includegraphics[width=.48\textwidth]{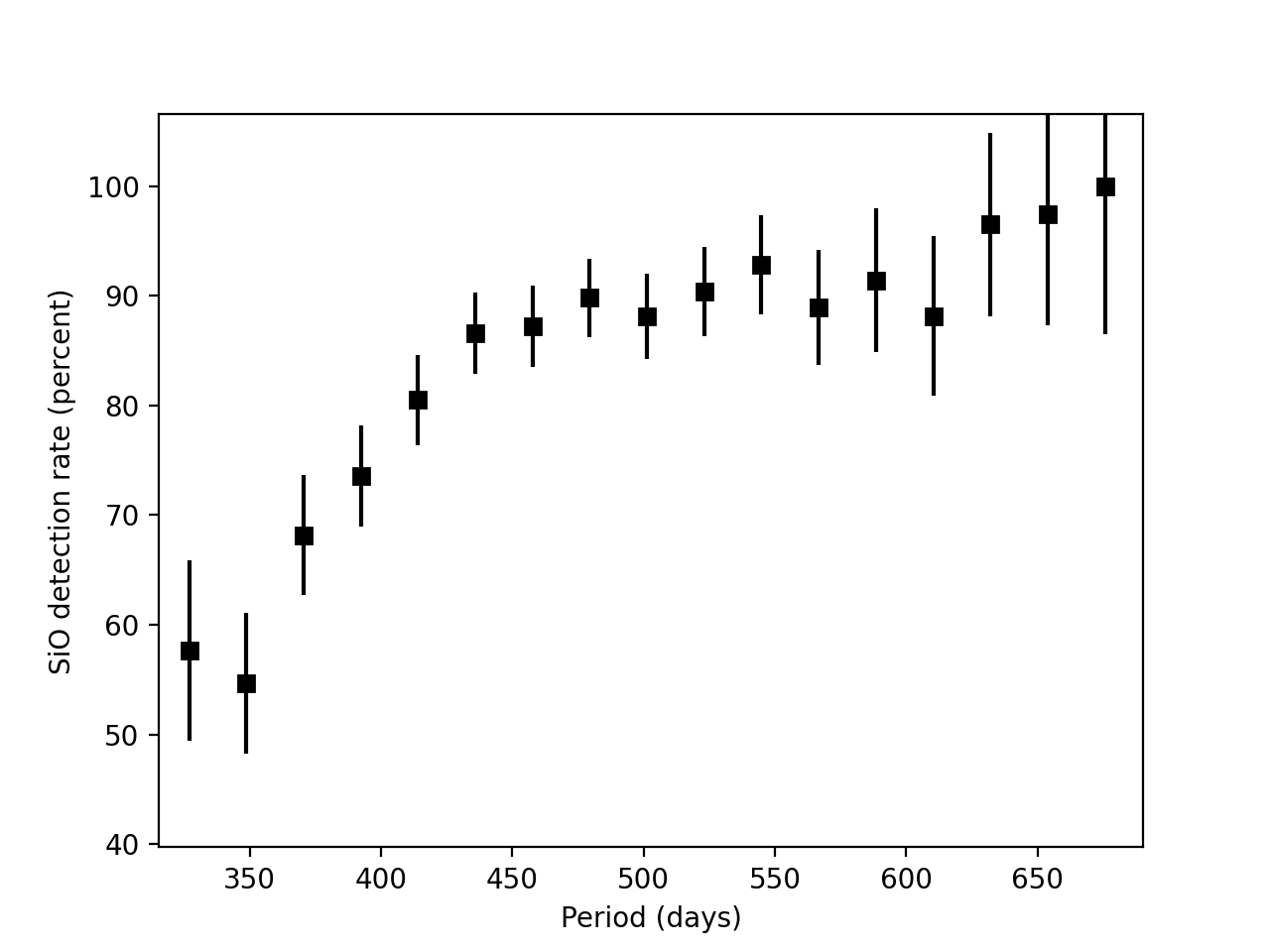}
    \caption{Period distribution of various samples. Top panel: Periods, as reported in the OGLE LPV catalog, for all sources identified as Mira variable by the OGLE team (light gray) as well as for sources that have been observed for SiO maser emission by the BAaDE survey (black dotted line showing sources with detections of these masers and solid showing sources without a detection). Also shown are the periods of the GC sample (dark gray). The number of sources in each bin for BAaDE samples and GC sample have been multiplied by 10 and 20, respectively, for readability. Bottom panel: SiO maser detection rate as a function of period derived from the bins in the top panel. Results are shown over the period range with good statistics ($\sim$350-650 days). Error bars are determined by the number of sources used to calculate the detection rate and go as $1/\sqrt{N_{\rm sources}}$.}
    \label{fig:periods}
\end{figure}

The full OGLE bulge catalog of Mira variables includes sources with periods between roughly 100 and 800 days with a peak around 310 days \citep{soszynski13, iwanek22}. The OGLE Mira sources that were also observed as part of the BAaDE project have periods on the longer side of this distribution, peaking near 430 days (see Fig.\,\ref{fig:periods}). 
BAaDE sources were selected from the MIR MSX catalog \citep{scc09}, where redder colors are known to roughly correspond to longer-period Mira stars, and so it is reasonable that BAaDE sources should correspond with relatively long-period OGLE Mira sources. 
This agrees with the $\sim$500-day average period derived for a sample of MIR-selected candidate Miras monitored by \cite{urago20}.

\par Additionally, the sources from the BAaDE catalog with detected maser emission have, on average, longer periods than the sample of nondetections. The dependence of SiO maser detection rate on period has been reported before \citep{imai02, deguchi04}, and this dependence is also evident in our sample (see right panel of Fig.\,\ref{fig:periods}).
We find that maser-bearing sources within the cross-matched BAaDE-OGLE sample have periods that peak around around 470 days whereas the sources with no maser detections peak around 390 days. 
Although both distributions have a large period range, the offset in their peaks is significant (Fig.\,\ref{fig:periods}).
This suggests that period could be used to optimize detection rate in SiO maser surveys. The SiO detection rate is much higher for Miras with periods over 475 days (90\%) than for Miras with shorter periods (79\%). 
\par The sample that is the main focus of this work, the GC sample, has an average period of around 470 days and their distribution peaks around 450 days. 
Thus these GC sources have long periods with respect to Mira variables in general, but are consistent with other maser-detected Miras. 
The distribution of the periods of these sources is likely a consequence of many competing factors including the maser-bearing nature of these sources, the metallicity and age of Miras in the GC, and the selection functions of the OGLE LPV and BAaDE catalogs. 
\par With this tendency toward long periods in mind, we note that many Mira PLRs become more scattered at periods of about 450 days and longer, likely due to a combination of hot-bottom burning and CSE effects \citep{whitelock03, yuan17b}. 
Therefore, it is important to consider these, and any other, effects that may drive uncertainty and scatter at the long-period end of Mira PLRs when considering the GC sample.  

\subsection{Reddening and extinction}
The intrinsic IR luminosities of the GC sample of Miras are expected to follow a PLR.
In this section, we define PL NIR magnitudes for a Mira with a certain period based on period-magnitude relations in the literature, and compare these PL magnitudes with those from the 2MASS survey to derive extinction values for the GC sources. 
\par To study the J-, H-, and K$_s$-band extinction of this sample we choose to use the relations from \cite{yuan17a}, where LMC Miras, including those with periods over 400 days, are shown to follow a quadratic relation between period and NIR magnitudes.
We note that \cite{yuan17b} provide similar relations but make corrections for stellar phase giving formally lower scatter; however, we use the values from \cite{yuan17a} because they are based on a sample with a higher portion of long-period sources. 
For example, only ten O-rich sources with a period over 500 days are used in the \cite{yuan17b} determination of the H-band relation and five are marked as outliers, whereas over 50 sources with periods over 500 days were included in \cite{yuan17a} for the H band.
Further, we have completed all of the analysis described in this paper using both sets of relations, and find that the \cite{yuan17b} parameters lead to the unphysical case of non-zero H-band extinction when K$_s$-band extinction is calculated to be zero.
We propose that the \cite{yuan17a} values are more appropriate for long-period sources, especially in H band, despite not being corrected for stellar phase.
From these relations we find J$^{\rm{PL}}$, H$^{\rm{PL}}$, and K$_s^{\rm{PL}}$ magnitudes for a Mira with a period, P in days:
\begin{equation}\label{plrJ}
J^{\rm{PL}}=-5.78 - (3.16(\log(P)-2.3)) - (3.07(\log (P)-2.3)^2)  
\end{equation} 
\begin{equation}\label{plrH}
H^{\rm{PL}}=-6.62 - (3.42(\log (P)-2.3)) - (2.85(\log (P)-2.3)^2)  
\end{equation} 
\begin{equation}\label{plrK}
K_s^{\rm{PL}}=-6.96 - (3.72(\log (P)-2.3)) - (2.75(\log (P)-2.3)^2) . 
\end{equation} 

We note that these relations are quadratic as opposed to linear. The quadratic nature is physically motivated by the hot-bottom burning which takes place in AGB sources with long periods, essentially giving a magnitude boost to these longer-period sources \citep{whitelock03, yuan17a, yuan17b}.
Using nonlinear relations such as these is critical for our sample because  we consider mostly sources with periods over 400 days (Fig.\,\ref{fig:periods}).
\par
\begin{figure}[t]
    \centering
    \includegraphics[width=.38\textwidth]{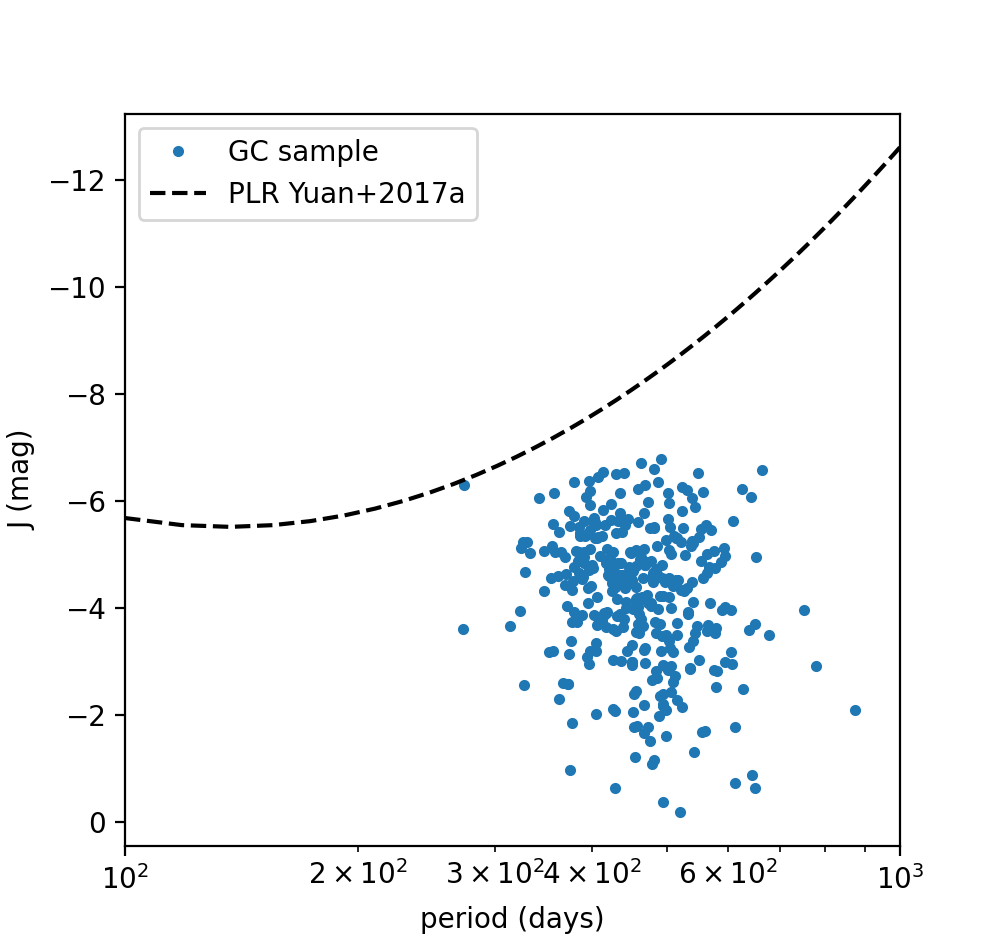}
    \includegraphics[width=.38\textwidth]{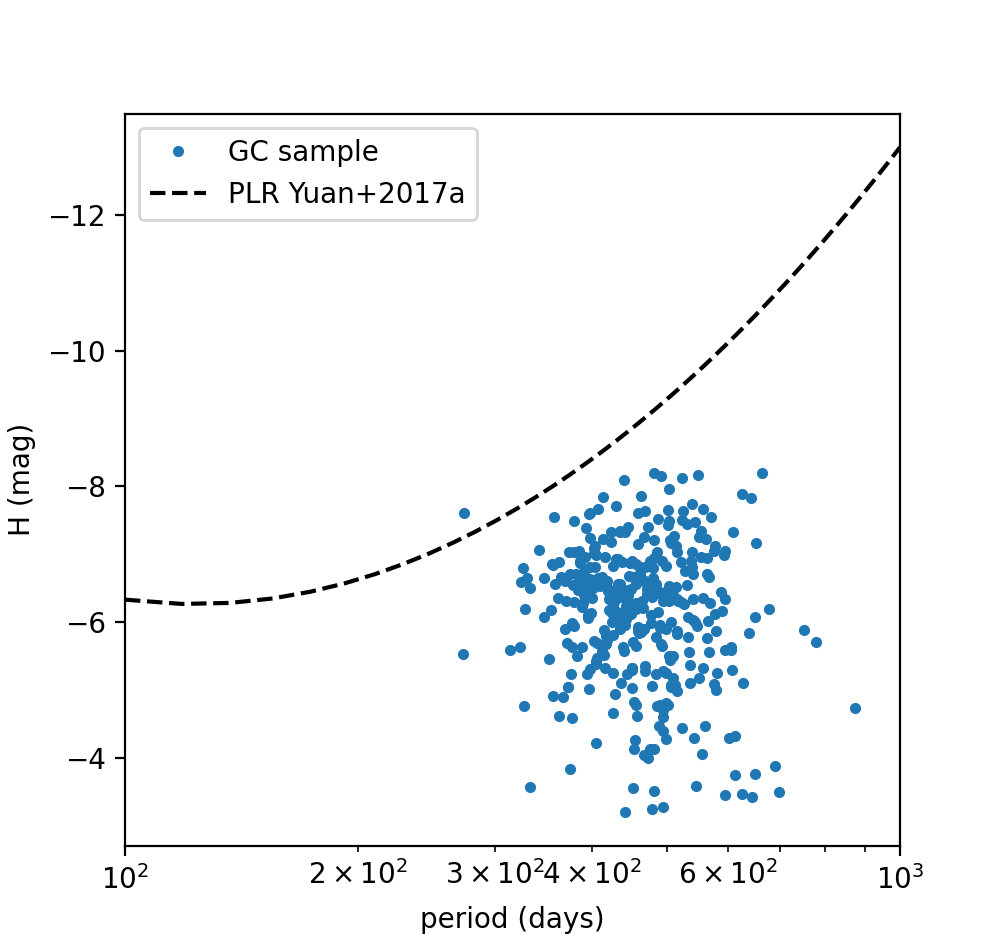}
    \includegraphics[width=.38\textwidth]{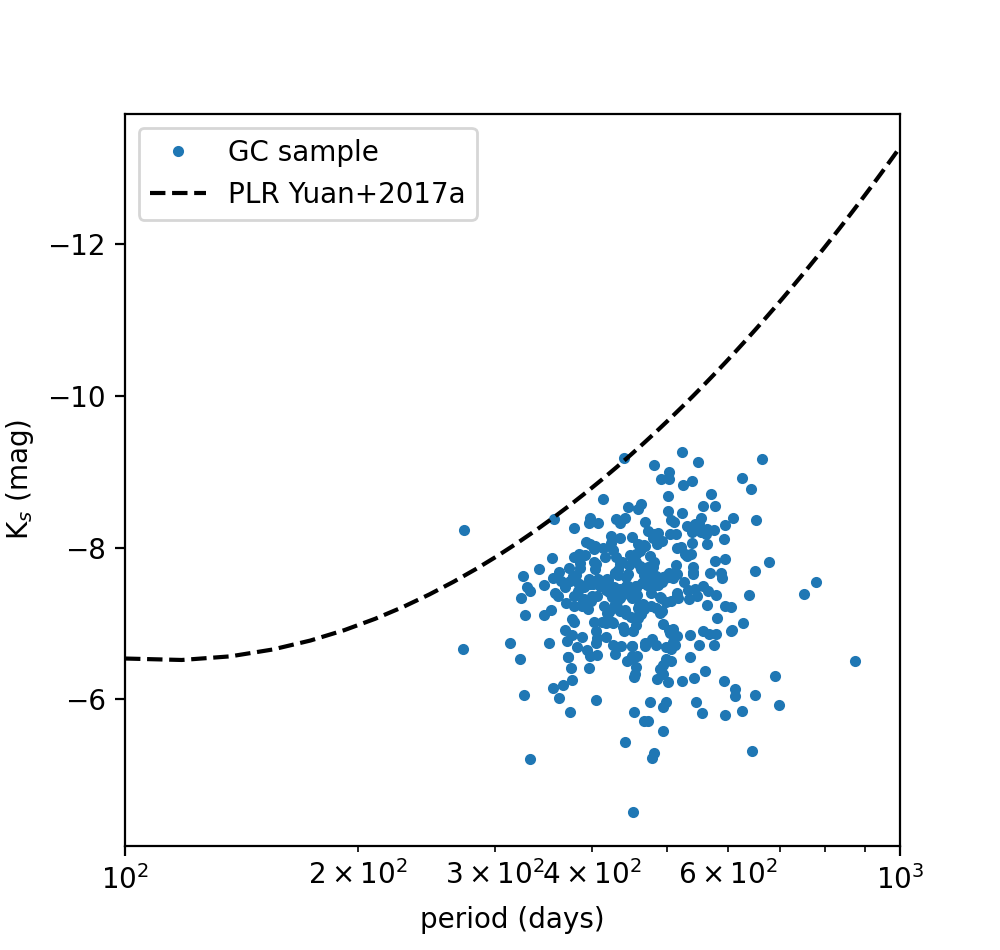}
    \caption{Uncorrected absolute 2MASS J, H, and K$_s$ magnitudes and periods of the GC sample. From this we interpret that extinction is still a large problem for NIR wavelengths causing NIR magnitudes to be fainter than the \cite{yuan17a} relations which we assume hold here. }
    \label{fig:periodK}
\end{figure}

 The periods and observed J-, H-, and K$_s$-band magnitudes of the 370 GC sources do not fit the above relations for Mira variables well, and the lack of fit can be explained by extinction---which is expected to be severe in the sample. 
Fig. \ref{fig:periodK} shows this lack of fit for the J, H, and K$_s$ bands, where it is clear that nearly all of the observed magnitudes are fainter than is predicted by the period-magnitude relation from \cite{yuan17a}.
Fainter magnitudes are indeed consistent with extinction being the main driver of the lack of fit, as is the increase in offset as a function of wavelength.
Several magnitudes of K$_s$-band extinction are well motivated for a sample of Mira variables in the GC which host masers (i.e., have at least some sort of CSE in which such masers can form). 
Several other samples of Miras from OGLE \citep{ita11}, and MACHO and SAAO data \citep{qin18} have shown that longer period Mira variables deviate significantly from PLRs and period-color relations even when corrected for interstellar reddening because of their thick CSEs.
Thus, while these heavily obscured sources will not contribute to refining the Mira PLR in NIR bands nor the distance to the GC, they can reveal information about the properties of reddening toward the Galactic center and in their own envelopes.
\par Comparing the J$^{\rm{PL}}$, H$^{\rm{PL}}$, and K$_s^{\rm{PL}}$ magnitudes calculated for these Miras to the J, H, and K$_s$ magnitudes observed by 2MASS, we can attribute the difference to the total (interstellar and circumstellar) extinction in those bands (e.g., $K_s - K_s^{\rm{PL}} = A_k$). 
This represents the approximate extinction caused by both interstellar and circumstellar material.
\par  
Table \ref{tab:ext} presents the extinction values, derived from periods, for each individual source, where the extinction has been split into interstellar and circumstellar contributions via the method in the following section.
\begin{table}
    \caption{GC sample extinction.}
    \centering
    \begin{tabular}{|c|c|c|c|c|c|c|}
    BAaDE & \multicolumn{3}{c|}{interstellar extinction} & \multicolumn{3}{c|}{circumstellar extinction}  \\
    name & J & H & K$_s$ & J & H & K$_s$ \\
ad3a-04238 & 2.61 & 1.52 & 0.91 & 0.88 & 0.31 & $-$0.13 \\
ad3a-04347 & 2.60 & 1.51 & 0.91 & 1.85 & 1.52 & 1.19 \\
ad3a-04350 & 0.86 & 0.50 & 0.30 & 6.04 & 4.13 & 2.64 \\
ad3a-04376 & 3.19 & 1.85 & 1.11 & 1.75 & 1.59 & 1.32 \\
ad3a-04404 & 0.65 & 0.38 & 0.23 & 3.32 & 2.28 & 1.57 \\
ad3a-04412 & 1.75 & 1.02 & 0.61 & 2.89 & .. & .. \\
ad3a-04415 & 3.12 & 1.81 & 1.09 & 0.16 & 0.05 & $-$0.05 \\
ad3a-04418 & 2.82 & 1.64 & 0.99 & $-$0.05 & 0.08 & $-$0.03 \\
ad3a-04425 & 2.01 & 1.16 & 0.70 & 2.30 & 1.90 & 1.23 \\
ad3a-04433 & 1.10 & 0.64 & 0.38 & 1.79 & 1.47 & 1.20 \\
    \end{tabular}
    \tablefoot{Calculated interstellar and circumstellar extinctions in the J, H, and K$_s$ bands (in magnitudes) for the GC sources. The typical errors on these values are discussed in Sect.\,\ref{sect:error}. The full table is available in the online journal. The first ten rows are presented here to demonstrate the form and content.  }
    \label{tab:ext}
\end{table}

\subsubsection{NIR interstellar extinction corrections} 
\par  We correct for interstellar extinction by identifying the location of the Red-Giant branch (RGB) in 2MASS data surrounding the coordinates of each of our GC sources. This means we exclusively use the 2MASS data to calculate interstellar extinction values with no need for period or maser information from the GC sources. For each of the GC BAaDE/OGLE sources in our sample we use the 2MASS sources that are within 4$^{\prime}$ to construct a J$-$K$_s$ versus K$_s$ color-magnitude diagram, and then fit a linear RGB to those sources similar to the methodology of \cite{messineo05}. This radius of 4$^{\prime}$ was chosen such that the area around the target source is small but still yields a robust RGB of about 600-1,000 sources in the color-magnitude diagram. Assuming both the particular BAaDE source and the majority of the 2MASS sources are about 8 kpc away we can then calculate the reddening in the area on the sky surrounding each target source following the method presented in \cite{messineo05}. 
This method is particularly well suited to our situation, where the sources we are interested in making interstellar extinction estimates for are likely near the GC, and therefore they are at the same distance as the bulk of the 2MASS RGB sources.
We use the RGB of the globular cluster 47 Tuc as a template, where the RGB stars in that cluster fit the following relation:

\begin{equation}\label{47tuc}
    {(J-K_s)}_0=2.19(\pm 0.02)-0.125(\pm 0.002){K_s}_0
\end{equation}
Use of this cluster as a template is advocated by \cite{messineo05}, based on the results of \cite{valenti04} which show that the RGB of 47 Tuc has a similar color to the RGB in bulge globular clusters. Furthermore, bulge globular clusters have been shown to have similar characteristics as compared to bulge field stars, especially regarding the RGB (e.g.,\citealt{minniti95}).
In addition, 47 Tuc has been specifically shown to have a population similar to bulge field stars, albeit with slightly different average metallicity \citep{rich13}. 
As \cite{ivanov02} showed that the RGB slope can be dependent on metallicity, we take the cautious approach of fitting both slope and intercept in each of our fields to allow for the fact that the metallicity of our fields can be different than that of 47 Tuc.
The reddening is then given by subtracting the template relation of 47 Tuc, $(J-K_s)_0$, from the $(J-K_s)$ color relation determined from our fits to 2MASS color-magnitude diagrams. This yields color differences over a range of K$_s$-magnitudes. We then select the median of the data set as the E(J$-$K) reddening, and convert this value to A$_K$ extinction using the following equation:
\begin{equation}
A_{K_{s}}=C_{JK}\times E(J-K_s)
\end{equation}
where $C_{JK}$ is a constant which depends on the value of $\alpha$ in the power-law relating wavelength and extinction: A$_\lambda \propto \lambda^{-\alpha}$. The value used here for $C_{JK}$ is 0.537 which is consistent with $\alpha=1.9$ as recommended by \cite{messineo05}. This method is applied to each of the 370 fields surrounding the GC sources.

\par Examining the color-magnitude diagrams produced for this sample reveals that 2MASS sources fainter than $K_s=12$ mag are not RGB objects and hence should be discarded in the RGB fitting. Moreover, photometric errors are less than 0.04 mag when objects brighter than 12 mag are considered, and previous work has also shown that a magnitude cut at $K_s<12$ is the most efficient in terms of fitting the RGB \citep{dutra2003extinction}. Therefore, we employ this cut before fitting the RGB. In addition to this faint magnitude cut, we also restrict the much brighter foreground stars in our calculation by excluding sources with $K_s<6$ mag. It has been shown that stars with $K_s$-band magnitudes brighter than 6 are mostly bluer foreground stars and can bias the extinction calculation in that direction \citep{trapp2018sio}. This generally leaves about 35-50\% of the 2MASS sources in our 4$\arcmin$ area or about 600-1,000 sources in each of the 2MASS color-magnitude diagrams.
\par
\begin{figure}[t]
     \centering
     \includegraphics[width=0.44\textwidth]{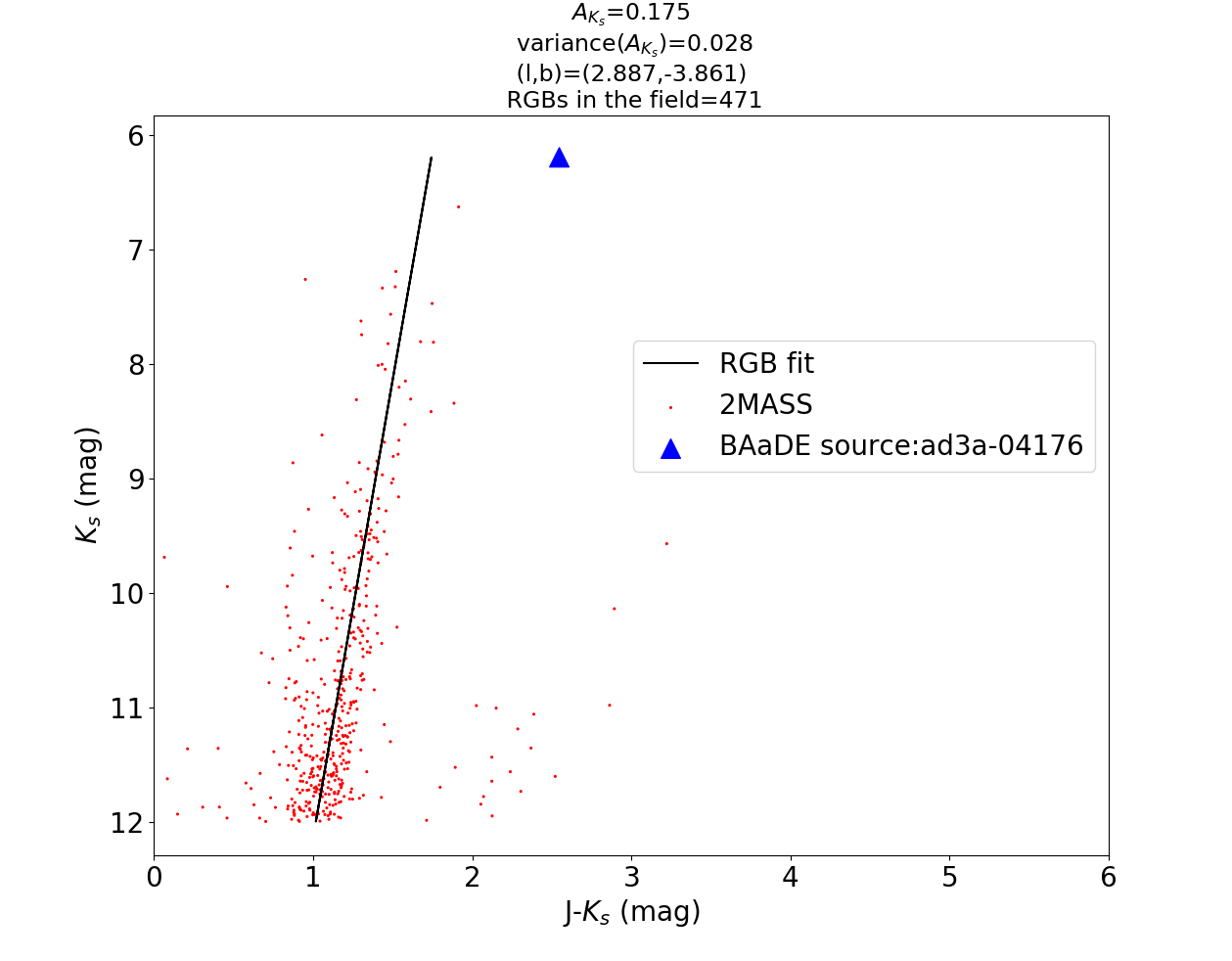}
     \includegraphics[width=0.44\textwidth]{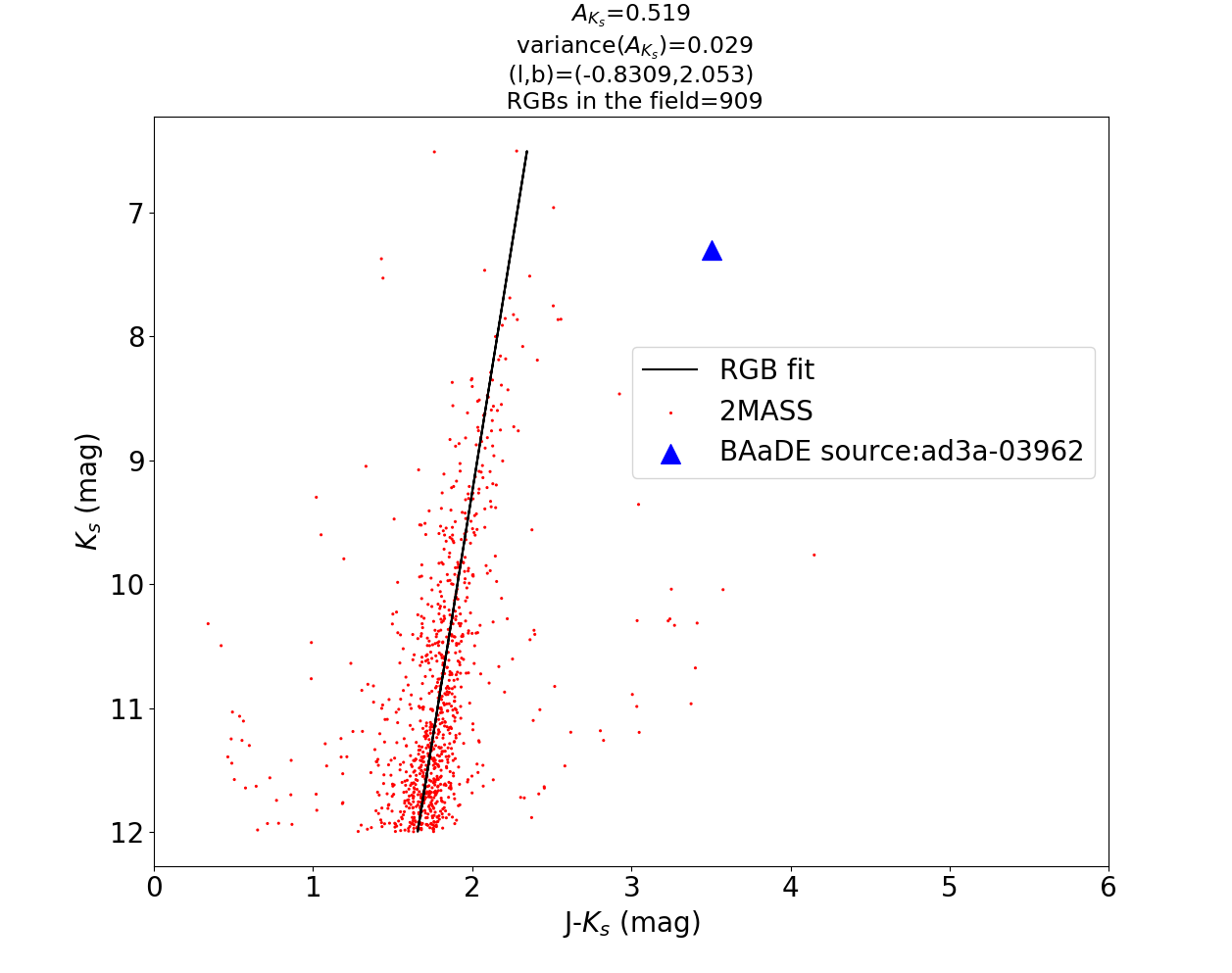}
    \includegraphics[width=0.44\textwidth]{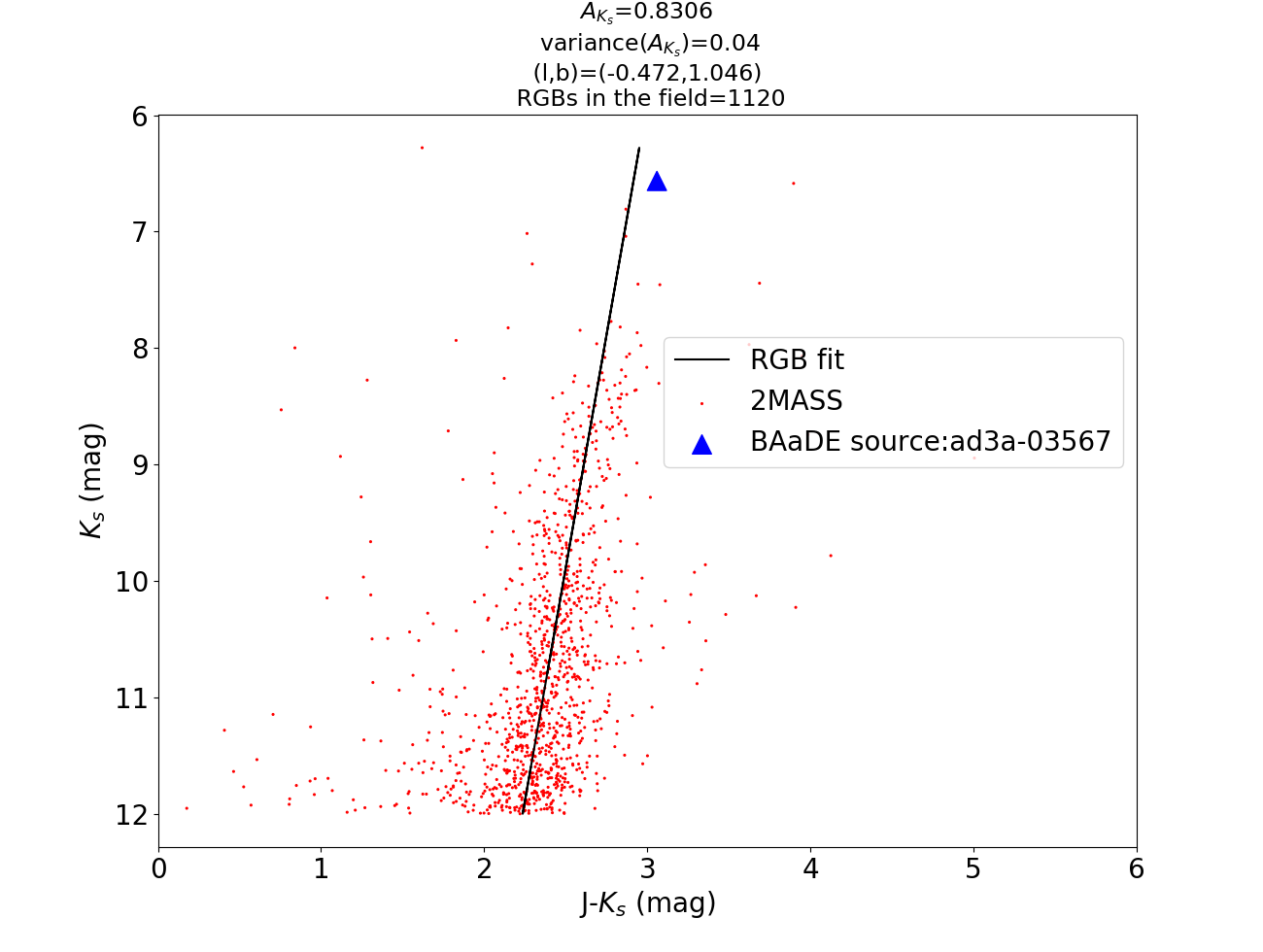}
     \caption{Each panel shows a color versus apparent magnitude plot of 2MASS sources (red dots) along with the much brighter BAaDE source (blue triangle) and our RGB fit (black straight line). All sources are within 4$^\prime$ of the coordinates shown at the top of each panel. The RGB fit is used to derive the reddening and extinction (A$_{K_s}$, shown at top of plot) as described in the text. This procedure was conducted for each source in the GC sample and these plots are shown as examples of a low-extinction field, and typical-extinction field, and a high-extinction field from top to bottom.}
     \label{fig:int_ext_J_K}
\end{figure}

 Once we apply the magnitude cuts to the 2MASS data, calculate the reddening by the above method, and shift the RGB by the calculated reddening, it is clear that our fit traces the RGB in the 2MASS data (Fig.\, \ref{fig:int_ext_J_K}). Additionally, we can put the BAaDE source on the same plot, and as expected this source, which should be on the AGB, is usually much brighter than the 2MASS counterparts (mostly on the RGB) in the same region \citep{messineo05}.
This is illustrated for three of the 4$\arcmin$ fields in Fig.\,\ref{fig:int_ext_J_K} where each panel shows the RGB fit, 2MASS data, and the BAaDE source, clearly indicating that the BAaDE sources are brighter and redder than most of their 2MASS counterparts, as expected. The variance has also been calculated to quantify the reliability of the fit, thereby reflecting the accuracy of the calculated extinction values; this is also used to quantify uncertainty in Sect.\,\ref{sect:error}. 
From these interstellar $A_{K_s}$ values we also calculate interstellar $A_J$ and $A_H$ using ratios of $A_J$/$A_{K_s}$ = 2.86 and $A_H$/$A_{K_s}$ = 1.66 \citep{messineo05}.
\par We compared the $A_k$ values obtained via this method to the maps of \cite{marshall06}, \cite{nidever12}, \cite{gonzales12}, and \cite{gonzales18}, and find generally very good agreement. For up to 0.5 mag of extinction, our values generally match these other maps very closely with no systematic offsets. Between 0.5 and 1.5 mag of extinction, we find that our values are often slightly lower (0.2 mag or less). This difference is likely caused by the fact that our values are specifically tailored to the exact fields in which we are interested. These fields are very near, but always just outside of, large dust features in the Galactic plane. This is a feature of the cross-match to OGLE which requires high quality I-band data to be observable, and thus sources cannot be fully optically obscured. Likely our values are more appropriate for the exact location of our sources while other catalogs average in the effects of the nearby dust for these fields. 

\subsubsection{Circumstellar extinction}\label{sect:csext}
\begin{figure*}
    \centering
    \includegraphics[width=.95\textwidth]{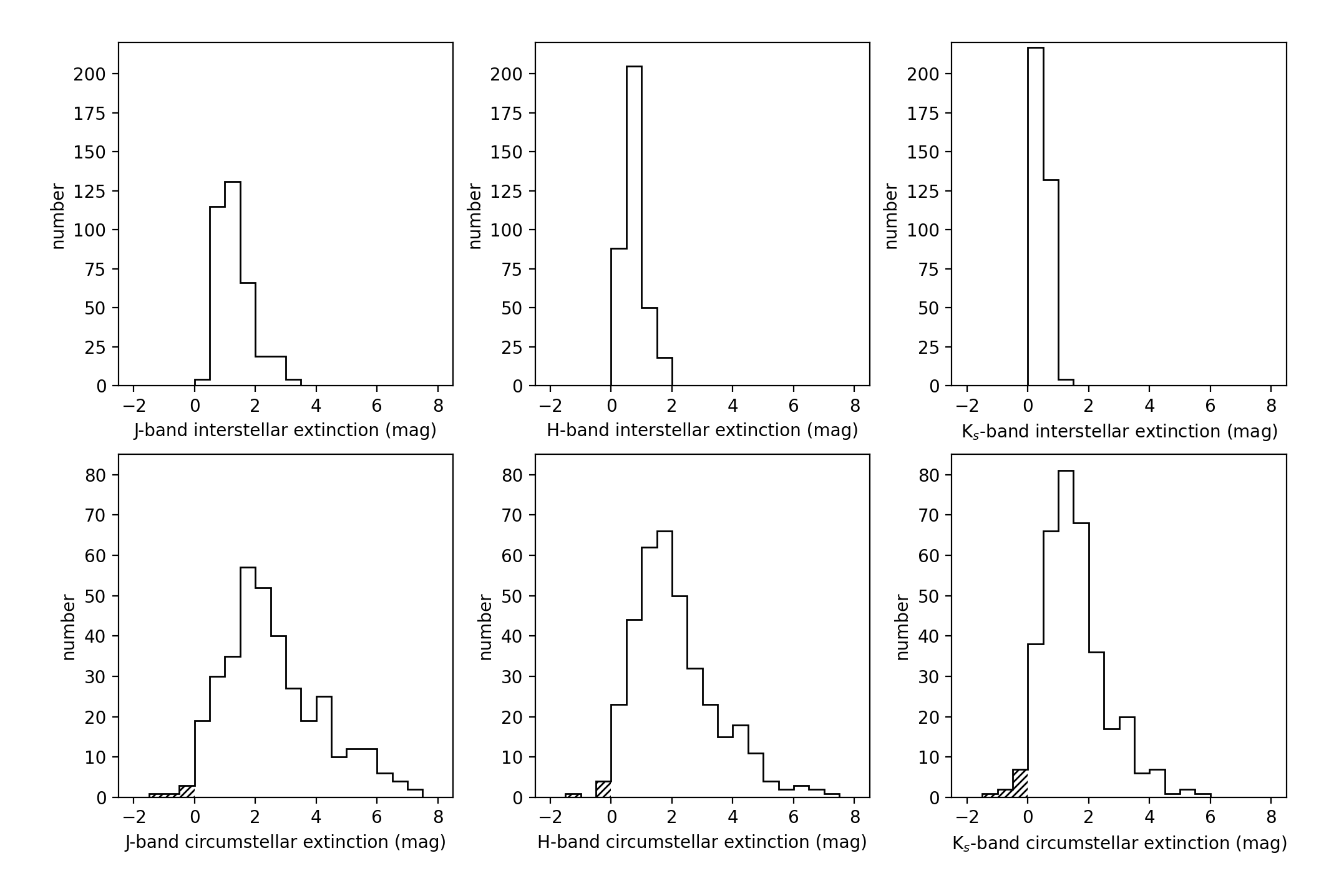}
    \caption{Separate interstellar and circumstellar extinctions as derived from period magnitude relations and our RGB-based interstellar extinction calculations. Our methods give unphysical (i.e., negative) circumstellar extinction values for about 3\% of the sample in the K$_s$-band (and 1\% in J and H), marked in the hashed region of the histogram. Unphysical values could be caused by our distance assumption being incorrect and/or by the fact that we use single-epoch magnitudes instead of averages. The high-end tails of the circumstellar distributions are likely affected by the same issues. However, on a population scale, our methods give an idea of the range and extent of circumstellar extinction.}
    \label{fig:csexthist}
\end{figure*}

Interstellar extinction for these sources ranges from roughly 0.2 to 0.9 mag in the K$_s$ band, meaning that between about 0 and 4 mag of extinction in the K$_s$ band remain unaccounted for and can be attributed to circumstellar extinction (see Table \ref{tab:ext} or Fig.\,\ref{fig:csexthist}). This points to a large spread in CSE extinctions even within this relatively homogeneous sample.
This spread can be seen in the last panel of Fig.\,\ref{fig:csexthist} for the K$_s$-band. 
\cite{ita11} report a similar range of circumstellar extinctions (large spread and up to 4 mag in the K$_s$-band) in LMC Miras, although fewer of their O-rich sources reach the high-extinction end of this distribution as compared to the GC sample. They show that circumstellar extinction values of 2-4 mag are reasonable by estimating that these extinctions are associated with mass-loss rates of roughly 10$^{-5}$ M$_\odot$ y $^{-1}$, using equations relating extinction and mass loss from \cite{vanloon07}.
\par Additionally, we note some of these derived circumstellar extinction values are negative and therefore unphysical.
These values could be caused by our distance assumption being incorrect or by the single-epoch 2MASS magnitudes which do not necessarily reflect the average magnitude over a stellar cycle. 
This problem only affects 3\% of our sample in the K$_s$ band (and 1\% in J and H, where all sources with negative J or H values also have negative K$_s$ values). All of the sources with negative extinctions have relatively short periods ($\lesssim$500 days) and relatively blue colors (J$-$K$_s<$ 4 mag). Not much CSE extinction was expected for these blue, short-period sources and statistical issues could have pushed the extinction into the negative. We find no specific region of the sky in which negative-extinction preferentially occurs. Even though individual extinctions may be negative, that is, not entirely accurate, by applying our methods to a large sample of sources, we can effectively probe the range and extent of circumstellar extinction on a population scale. Likewise the highest extinction values (i.e., those over 4-5 mag in K$_s$) are likely affected by our distance assumption and/or stellar phase. 
On the whole, we see the circumstellar extinction for our long-period Mira sample can easily be more severe than interstellar extinction toward the GC in the NIR. 
\par Extinction across various bands is often expressed as the total-to-selective extinction ratio, $C_{JK} = \frac{A_K}{E(J-K)}$, and we find that this value also varies widely for the circumstellar extinction in the GC sources. Most of our circumstellar $C_{JK}$ values lie between 0 and 2.0 for our sample with an average value of $\sim$1.1; whereas values of $\sim$0.5 are typical of the interstellar medium in the literature \citep{nishiyama09, majaess16, dekany19}. 
This shows not only that the amount of extinction among the CSEs in this sample is highly variable, but also that the reddening law cannot strictly be expressed as $C_{JK} = 0.5$ or any similar single value. We find that $E{(J-K)}$ can be quite close to zero in our sample, and that consequently small changes in this value cause large spread in $C_{JK}$. Therefore, we explore the reddening law in this sample more in Sect.\,\ref{disc:reddeninglaw} using a different parameterization.  
\subsubsection{Error budget of circumstellar extinction values}\label{sect:error}
Uncertainty in the calculated circumstellar extinctions is caused by a combination of photometric uncertainty, variability, distance assumptions, uncertainty in Equations \ref{plrJ}-\ref{plrK}, and our interstellar calculations. Here we make estimations to quantify the contribution of each of these sources of uncertainty. 
\par One of the smallest contributions to this budget are the 2MASS photometric uncertainties, which are typically estimated to be accurate to within $\sim$0.025 mag at the time of observation. 
However, these are single-epoch measurements and how these relate to the average K$_s$ magnitudes is a large source of uncertainty.
We have examined the feasibility of tracing the OGLE lightcurves back to the 2MASS epoch in order to estimate the I-band phase of each source during the 2MASS observation. 
Unfortunately, for most sources, more than 6-7 full stellar cycles have passed between the 2MASS epoch and the first epochs of well-sampled OGLE data, and tracing the lightcurves back over such a long period of time adds as much uncertainty as it corrects for. 
However, we can use this method to estimate the uncertainty caused by variability on a population scale. 
To estimate the effects of variability in the K$_s$ band, we first transform the I-band light curve to one in the K$_s$ band using a phase-shift of 0.08051 and an amplitude ratio of 0.368 \citep{iwanek21}; for J and H bands we use phase shifts of 0.02818 and 0.05430, and amplitude ratios of 0.613 and 0.463 respectively.  
The 2MASS photometric point is used to anchor this transformed lightcurve, and then an average magnitude is extracted, similar to the method used by \cite{sun22}.
Average magnitudes calculated in this way are very rough given the large amount of time between the 2MASS and OGLE photometry (often ten years), and we note again that we do not use these average magnitudes in our extinction estimates. 
Instead, we examine the differences between the average and the 2MASS magnitudes (K$_s^{\rm{2MASS}}$-K$_s^{\rm{ave}}$) for all of the sources in our sample in order to quantify the effects of variability on the 2MASS measurements in the sample as a whole.
Within our sample, these differences average around 0.004 mag and have a standard deviation of 0.37 mag for the K$_s$ band. For the J band this yields $\pm$0.62 mag; for the H band $\pm$0.46 mag.
We take these values as the uncertainty caused by variability. Note that this does not account for long term trends where the average magnitude drifts  over the course of many years or many stellar cycles. 
\par
\begin{figure}
    \centering
    \includegraphics[width=0.45\textwidth]{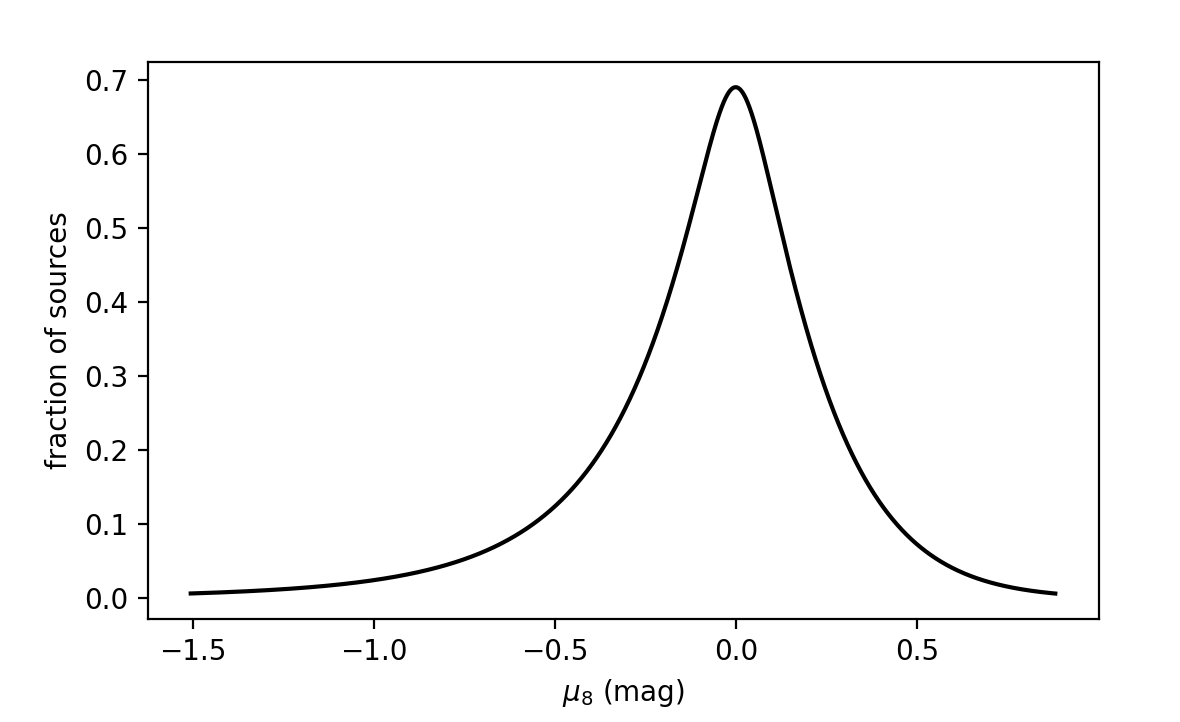}
    \caption{Magnitude error ($\mu_8$) incurred by assuming all sources are at a distance of 8 kpc if instead the sources are centered at 8 kpc and distributed exponentially in Galactic radius.}
    \label{fig:mu}
\end{figure}

 To address the uncertainty caused by our assumed distance of 8 kpc we first note that the GC sample should consist almost entirely of bulge sources due to their high velocities and the metallicity required to host SiO masers. 
We employ the bulge AGB distribution discussed in \cite{jackson02}, where the number of AGB sources as a function of Galactic radius, $r$, goes as $e^{-r/0.8\,\rm{kpc}}$. To relate this distribution to uncertainties in magnitude we use the distance modulus formula normalized to the GC distance:
\begin{equation}  \label{distmod}
    \mu_8=5 \log{ \frac{d}{8 \rm{kpc}}}
\end{equation}
where $\mu_8$ is the magnitude error introduced by assuming a source is at a distance of 8 kpc when the actual distance from the Sun is given by $d$. Combining these gives the fraction of bulge sources expected for each value of $\mu_8$ as shown in Fig.\,\ref{fig:mu}. We note that this calculation included the full 3D relation between $r$ and $d$ (the distance from the GC and the distance from the Sun, respectively) but only allowed for sources within our sampled latitude and longitude range.
Integrating the number of sources as a function of $\mu_8$ yields that $\sim$70\% (i.e., similar to one standard deviation for a Gaussian distribution) of sources are expected to have $\mu_8$ values between $-$0.28 and 0.25 mag. 
Thus, we roughly approximate 0.28 mag as the uncertainty caused by distance in our sample. Shifting the GC distance by 0.2 kpc causes a shift of $\sim$0.05 mag in $\mu_8$ which is much less than our estimated uncertainty.
\par The uncertainty in our interstellar extinction values is calculated by propagating the uncertainty of our RGB fits with the uncertainty in the fit to the template RGB of 47 Tuc (0.04 mag, see Eq.\,\eqref{47tuc}) to obtain an uncertainty in $E(J-K_s)$.  This reddening uncertainty is then propagated with the uncertainty of $C_{JK}$ for a final value associated with interstellar extinction. 
The uncertainty of the value $C_{JK}$ is taken to be 0.1, which we find to be a good estimate based on the values presented in Table 1 of \cite{messineo05}. 
The final value of interstellar uncertainty changes from source to source as the value depends on the quality of the RGB fit in the field around each source; values range from 0.19 to 0.27 mag with $\sim$70\% of the sources having uncertainties of 0.21 mag or smaller. 
\par Finally, the uncertainties in the period-magnitude relations used to derive J$^{\rm{PL}}$, H$^{\rm{PL}}$, and K$_s^{\rm{PL}}$, are taken as the scatter given in \cite{yuan17a}. These are 0.27, 0.27, and 0.24 mag, respectively. 
\par Adding these five sources of uncertainty in quadrature for each band gives circumstellar uncertainties of 0.76, 0.64, and 0.56 mag for the J, H, and K$_s$ bands, respectively. 
This exercise shows that although we are far from precision circumstellar extinction calculations, 3-4 mag of spread in these values cannot be accounted for purely by uncertainty.
The large spread in values is in part physical. 
Further, the lack of precise and accurate distance estimates, interstellar extinction, variability-related, and period-magnitude-relation uncertainties have an almost equally large impact, especially in the K$_s$ band. We note again that neither Gaia parallaxes nor the PLR-derived distance from \cite{iwanek23} are well suited to these distant stars near the plane with significant envelopes.

\subsection{Mid-infrared period-magnitude relations} 
Having demonstrated that NIR circumstellar extinction can be large in our sample, we can investigate period-magnitude relations at longer, MIR wavelengths, where both interstellar and circumstellar extinction are less of an issue. 
Up to this point we have focused on existing J, H, and K$_s$ period-magnitude relations as a tool for estimating expected magnitudes, as it has been argued by many authors that these bands (and K$_s$ especially) are most suitable for producing relations with small scatter (e.g., \citealt{whitelock13}). 
However, it has also been shown that there is a difference between ``optical Miras" and ``infrared Miras"---the later still suffering from considerable circumstellar extinction in the K$_s$ band and often having no optical counterpart \citep{wood03, ita11}. The GC sources fall strongly into the IR Mira category and therefore still suffer from too much circumstellar extinction at NIR wavelengths to easily construct period-magnitude relations at these wavelengths (recall Fig.\ref{fig:periodK}). 
Therefore, we have also compiled WISE and MSX magnitudes for these sources and can explore period-magnitude relations across a range of wavelengths.
To investigate these longer wavelengths we correct for interstellar extinction, where the K$_s$-band interstellar extinctions from Table \ref{tab:ext} have been transformed to each band using $\alpha=-1.9$ \citep{messineo05}.

\par
\begin{figure*}[h!]
    \centering
    \includegraphics[width=.9\textwidth]{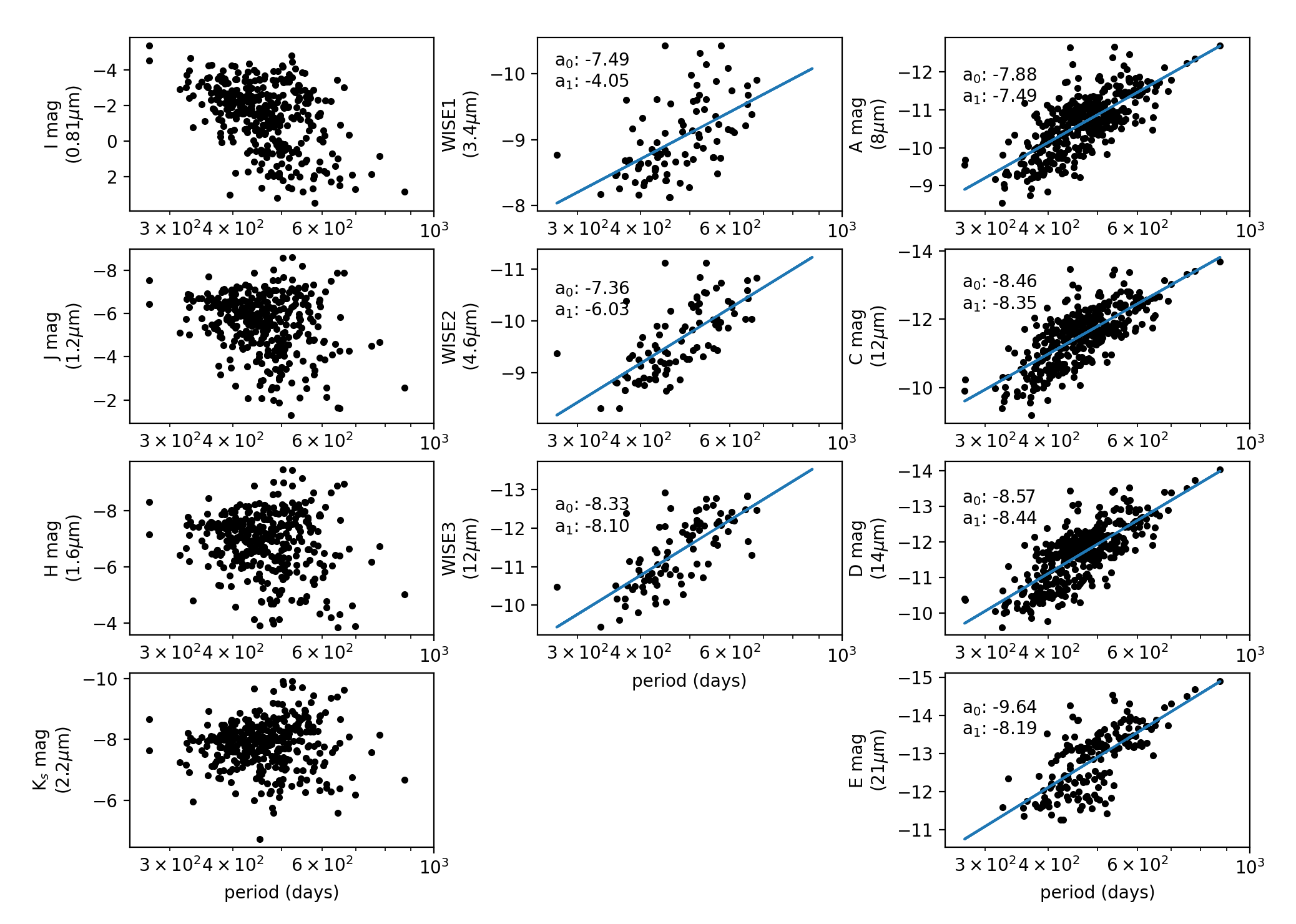}
    \caption{Period versus absolute magnitude plots for wavelengths from 0.9 to 21 $\mu$m. The y-axis in each plot is given in absolute magnitudes, where the central wavelengths of the I, J, WISE1, WISE2, WISE3, A, C, D, and E bands are 0.81, 1.25, 3.4, 4.6, 12, 8.0, 12, 14, and 21 $\mu$m respectively. Fits and their corresponding parameters are shown for WISE and MSX bands, where the parameters, $a_0$ and $a_1$, are obtained by fitting to the equation $M=a_0+a_1(log(P)-2.3)$.}
    \label{fig:pls_lotomags}
\end{figure*}

 We show in Fig.\ref{fig:pls_lotomags} that MIR magnitudes show a correlation with period in this sample, but that there is still considerable scatter in these relationships. 
At around 3 or 4 $\mu$m the presumably-circumstellar-extinction-related effects that disrupt observable period-magnitude relations are less substantial, and positive correlations between extinction-corrected (interstellar only) magnitude and period can be seen. 
We fit linear relations of the form $M=a_0+a_1(log(P)-2.3)$ (where $M$ is the absolute magnitude) to the WISE and MSX bands, and the parameters, $a_0$ and $a_1$, resulting from these fits are shown on the corresponding figure panels. 
Although we have discussed the improvement that quadratic PLRs usually yield for samples with periods longer than 400 days, we find in our case there are too few periods shorter than 400 days to allow for a reliable quadratic fit. 
Thus we present only the parameters of linear fits. 
The scatter in these relations could be caused by our assumption that these sources are all at the exact same distance, by the limited-epoch IR measurements (not averaged over stellar cycle), by circumstellar effects, or could be intrinsic. 
Note that we made no attempt to correct for circumstellar extinction in these plots because converting circumstellar extinctions from 2MASS bands to other bands requires one to make assumptions about the circumstellar reddening law or $\alpha$ value. 
As we show in Sect.\,\ref{disc:reddeninglaw} our circumstellar extinctions do not appear to follow typical interstellar reddening laws and so making such assumptions would obscure our results in this case. 
We discuss the behavior of these diagrams more in Sect.\,\ref{sec:pldis}.

\section{Discussion}
Within our sample there is an observable correlation between magnitude and period for wavelengths between 3.4 and 21 $\mu$m. To further investigate these period-magnitude relations at longer wavelengths, we focus in particular on the WISE bands where we can compare to existing relations based on LMC data. Additionally, we discuss the relationships between our calculated extinctions in various NIR bands and what these relations imply for dust in the CSE.
\subsection{WISE-band period-magnitude relations and MIR excess}\label{sec:pldis}
We compare our results to those obtained for OGLE Miras in the LMC. 
In particular, \cite{iwanek21} have recently explored these relations including data for longer period ($>$ 400 days) Miras. 
\begin{figure*}[t]
    \centering
    \includegraphics[width=0.6\textwidth]{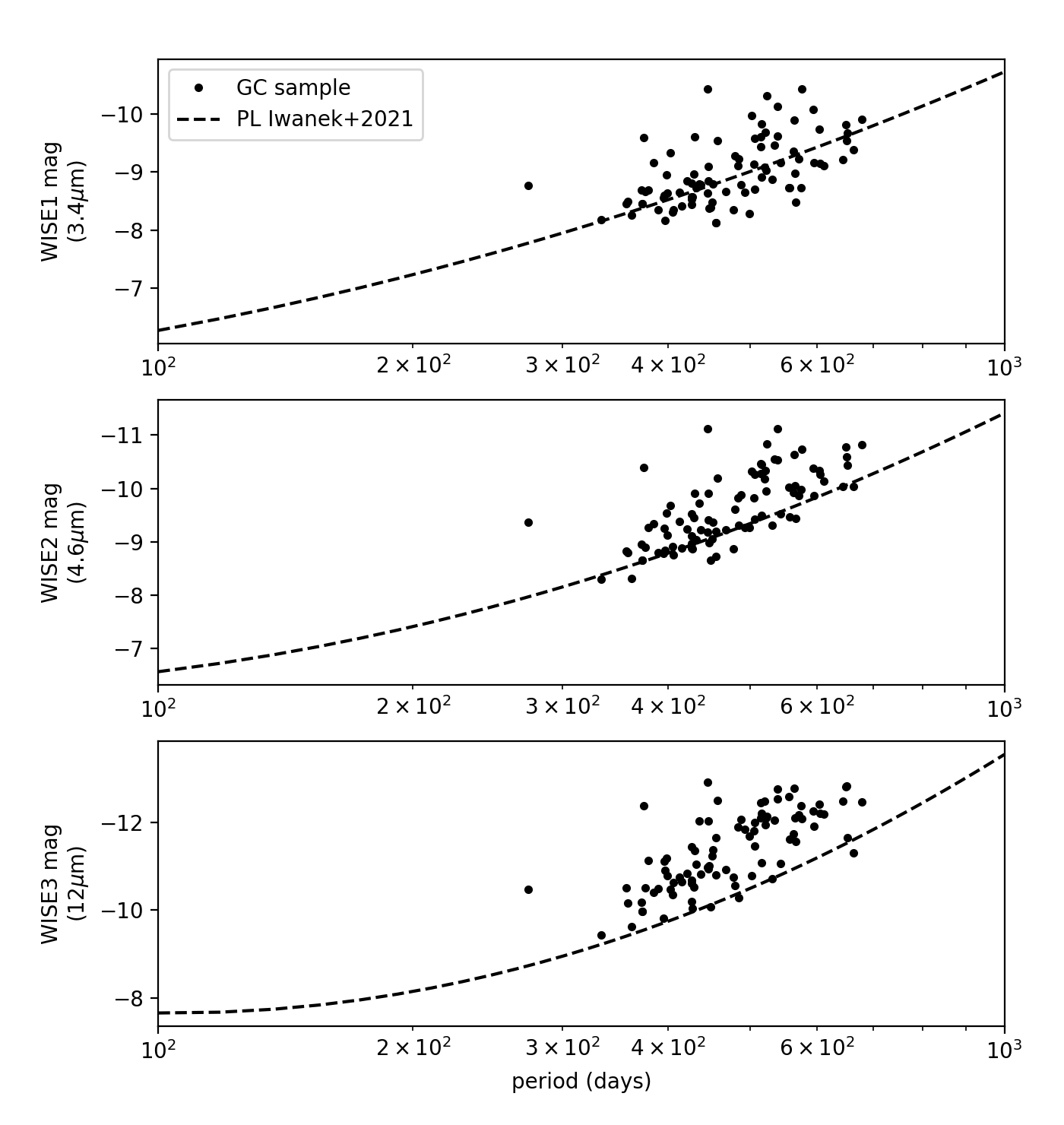}
    \caption{WISE magnitudes and periods of the GC sources as compared to period-magnitude relations derived from OGLE Miras in the LMC \citep{iwanek21}. An offset is seen between the data and the LMC relation---especially when considering the longest wavelengths---likely indicating that circumstellar emission becomes important at these wavelengths in the GC sample.}
    \label{fig:wisecompare}
\end{figure*}
In Fig.\ref{fig:wisecompare} we show our interstellar-extinction-corrected GC Mira magnitudes and periods as compared to the period-magnitude relations presented in \cite{iwanek21}. In general, our data show brighter Mira magnitudes at a given period than the LMC fits, and this differential widens for longer wavelengths. 
\par These brighter magnitudes could be the result of CSE effects. \cite{ita11} show that carbon-rich (C-rich) AGB sources with significant envelopes in the LMC lie below (are fainter than) the period-magnitude relations they derive for wavelengths less than 3 $\mu$m, indicating CSE extinction. However, they also show that the same sources with prevalent envelopes often lie above (brighter than) period-magnitude relations for wavelengths greater than 3 $\mu$m. This shows that the effects of circumstellar extinction become minimal around 3 $\mu$m, and may even be replaced by emission from the envelopes at higher wavelengths. Furthermore, \cite{urago20b} show that OZ Gem, a MW O-rich Mira with a substantial CSE, is positioned on a K-magnitude-period diagram among C-rich LMC Miras, suggesting that the effect seen in \cite{ita11} is related to CSE prevalence more than chemistry. 
\par In our sample, there is similar evidence for the weakened effect of circumstellar extinction in the WISE 1 band and at longer wavelengths; period-magnitude correlations are visible in WISE and MSX data but not shorter wavelengths. 
In addition, we see evidence for CSE contributions to the stellar brightness when comparing to \cite{iwanek21} for WISE 1-3 (Fig.\,\ref{fig:wisecompare}). 
For CSEs with temperatures $\lesssim$300 K, a simple blackbody approach would suggest that the circumstellar emission would be greatest in WISE band 3 followed by band 2 and then band 1. This could explain the increasing offset with increasing wavelength seen in Fig.\,\ref{fig:wisecompare}, but more detailed modeling accounting for optical depth is needed to explore this further.
\par Alternatively, the offset in our data and the LMC MIR relations could be initial evidence that period-magnitude relations in the MW and LMC are fundamentally different at these longer wavelengths, and that metallicity or other environmental factors affect these relationships. It is not likely that the offset is caused by a systematic problem with our distance assumption, because we would not expect the offset to be dependent on wavelength in that case. As the GC sample was selected from a MIR catalog, has been shown to have very long periods, hosts maser emission, and shows severe CSE extinction, we find it most likely that the sample is showcasing some of the most extreme effects of CSEs on period-magnitude relations, and that the offset is indeed caused by (or at least dominated by) CSE contributions to the WISE magnitudes, but we do not rule out environmental effects.

\subsection{Correlations of circumstellar extinctions across bands}\label{disc:reddeninglaw}
\begin{figure*}[t]
    \centering
    \includegraphics[width=0.95\textwidth]{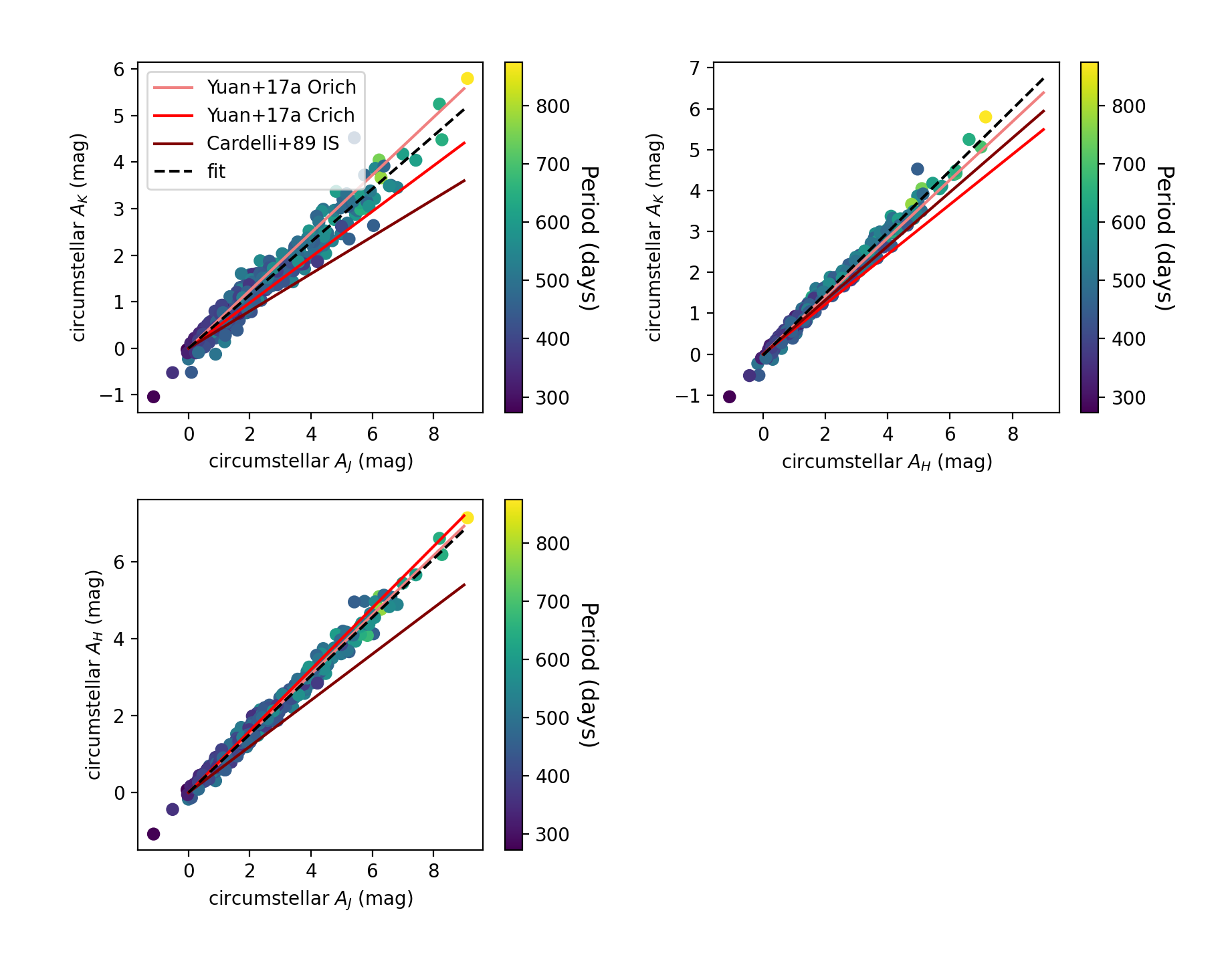}
    \caption{Colored dots show the circumstellar extinctions in the J, H and K$_s$-bands and their correlations with each other, where the color scale shows the periods of the sources in days. Colored lines show the relations between extinction in two bands as derived from the interstellar reddening law of Cardelli et al.\,(\citeyear{cardelli89}; maroon), the CSE extinction in O-rich LMC Miras as determined by Yuan et al.\,(\citeyear{yuan17b}; pink), and the CSE extinction relation in C-rich Miras from the same study (red). The fits to the data here (black dashed lines) generally match the relations for O-rich LMC Miras more closely than relations from the interstellar law. }
    \label{fig:a_jhk}
\end{figure*}

Calculated J, H, and K$_s$ circumstellar extinctions are strongly correlated as shown in Fig.\,\ref{fig:a_jhk}, and the slopes from linear fits to the circumstellar extinctions in each pair of bands are an expression of the reddening law in the CSE environments of the GC sample.
Specifically, the slopes presented are derived from fits with a fixed intercept of zero and exclude negative extinction points. This gives empirical values of A$_J$/A$_H$=1.30, A$_J$/A$_K$=1.75, and A$_H$/A$_K$=1.35 for the circumstellar environment. 
From Fig.\,\ref{fig:a_jhk} it is also clear that these extinctions are strong functions of period, as is expected for extinction caused by circumstellar effects. 

\par We compare the slopes of our circumstellar extinction data to interstellar extinction laws, as well as the circumstellar values from \cite{yuan17b}, who presented a similar comparison in their work. 
Similar to their study we find that our circumstellar extinctions do not match interstellar extinction laws well.
Specifically, we see relatively more extinction in the K$_s$-band than J- and H-bands compared to typical values assumed for interstellar extinction. 
Comparing just the J- and H-bands, we find more extinction in H as compared to interstellar reddening. 
These tendencies suggest that there are significantly more grains with sizes around $\sim$1\,$\mu$m in CSE environments than in interstellar environments, based on a preliminary grid of model SEDs of circumstellar envelopes of AGB stars, generated using the radiative transfer code DUSTY \citep{Ivezic2012} with different grain-size distributions. A standard dust grain size distribution following a power law with a maximum dust grain size of 0.25 $\mu$m (i.e., the default in DUSTY) does not reproduce our relation between K$_s$ and J-band circumstellar extinction; however the same distribution with a maximum grain size of 1 $\mu$m does reproduce this relation well.
Detailed modeling is required to further explore the physical reasons for these reddening differences, but is outside of the scope of this paper.
\par We do find that our slopes agree well with the slopes from \cite{yuan17b} for O-rich LMC Miras, with especially good agreement in the H - J circumstellar extinction plane. 
This is an indicator that NIR circumstellar reddening is similar for AGB sources in the LMC and the MW, at least in the long-period regime primarily sampled here. However, as mentioned, additional work remains to be done before precision circumstellar extinction values are achievable in the MW.
This agreement between the circumstellar extinction slopes for our sample and the O-rich LMC sample is expected because the GC sample also consists of O-rich Miras. 
In contrast, C-rich circumstellar environments have been shown to be very different from O-rich ones in terms of molecular content \citep{bujarrabal94}, dust properties \citep{devries10, messenger13}, color \citep{ortiz05, suh11, lewis20b}, and extinction properties.

\par It is inherently challenging to compare highly optically obscured (due to interstellar extinction) sources (i.e, the GC sample) with a less obscured sample (i.e., the LMC sample). Hence, it is notable that we find the clear result that the reddening over the 1-2 micron range between the circumstellar envelopes from these two samples is similar, especially given the very different metallicity environments in which these stars reside.

\subsection{Extinction correlations with color}
\begin{figure*}[t]
    \centering
   \includegraphics[width=1\textwidth]{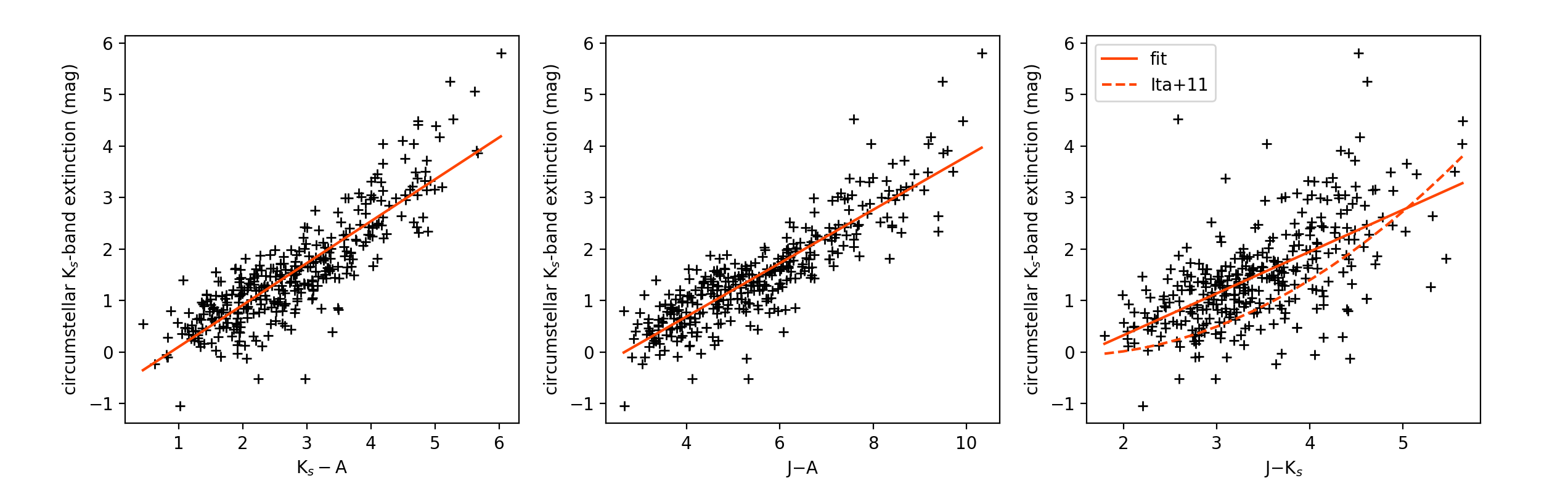}
    \caption{Correlation and fit for CSE K$_s$-band extinction, as calculated via our methods, and interstellar-reddening-corrected IR colors. }
    \label{fig:colorext}
\end{figure*}

Circumstellar extinction values in the NIR are also correlated with the observed colors of our sources. 
Specifically, there is a strong correlation between $A_K$ and K$_s-$A or J$-$A color (see Fig.\,\ref{fig:colorext}), where K$_s-$A and J$-$A colors are calculated using 2MASS and MSX magnitudes, that is, combining NIR and MIR data.
These correlations are unsurprising because they fit the very basic picture that thicker envelopes cause more extinction, are redder, and belong to Miras with longer periods.
However, the parameters of these empirical relations are still of interest because color can be relatively easily observed and can subsequently be used to estimate circumstellar extinction. 
The fits to these strongest color-extinction trends are presented as Eq.\,\ref{AkKA} and \ref{AkJA}, and we also present the A$_k$ versus J$-$K$_s$ trend as Eq.\,\ref{AkJK}:
\begin{equation}\label{AkKA}
    A_K = (0.81 \pm 0.077) (K_s-A) - (0.71 \pm 0.026)
\end{equation}
\begin{equation}\label{AkJA}
    A_K = (0.52 \pm 0.091) (J-A) - (1.39 \pm 0.016)
\end{equation}
\begin{equation}\label{AkJK}
    A_K = (0.81 \pm 0.195) (J-K_s) - (1.30 \pm 0.055),
\end{equation}
where the uncertainties shown are only the uncertainties of the final fit.
These relations are certainly not one-to-one as is evidenced in the scatters in the extinction in Fig.\,\ref{fig:colorext} (0.53, 0.48, and 0.78 mag for one standard deviation in K$_s-$A, J$-$A, and J$-$K, respectively), but we suggest that they can be used to curate samples of maser-bearing Mira stars which mitigate the effects of circumstellar extinction.
\par We also note that the relation between J$-$K$_s$ color and K$_s$-band extinction, although very scattered in our data, has been reported on for LMC Miras. We show a comparison between our linear fit and the second-order relation derived from the LMC Miras from \cite{ita11}. The relation from \cite{ita11} is consistent with our data, within the large amount of scatter. We note briefly that the fit from \cite{ita11} is driven mostly by the C-rich stars in their sample---with O-rich stars occupying only the bluest section of their diagram, but given the spread in our data we do not comment further.

\section{Conclusions}
\par Maser sources present a path forward for studying Mira period-magnitude relations in the MW, as distances to these stars can be measured via maser parallaxes (e.g., \citealt{sun22}). 
We explore circumstellar extinction and period-magnitude relations in a sample of SiO-maser-bearing sources near the GC in order to understand the biases that maser-bearing sources may introduce as compared to the general Mira population, as well as the effects of the high-metallicity environment of the Galactic bulge as compared to previously explored low-metallicity LMC populations.
\par About 7\% ($\sim$2,900) of the OGLE bulge Miras were searched for SiO maser emission in the BAaDE survey, and nearly 2,500 were detected. The detection rate of SiO masers among OGLE Miras is high at 84\%, and these sources show that the SiO detection rate among Miras is a strong function of period. Specifically, Miras with periods over 475 days have an even higher SiO-maser-detection rate of 90\%. Therefore, Miras that are also SiO-maser hosts are likely to have especially long periods, where we note that sources with such long periods require more complex (i.e., non-linear) PLRs to describe their behavior \citep{whitelock03}. 
\par Within these 2,500 maser-bearing Miras, we consider 370 sources to be likely in the GC based on their high line-of-sight velocities (absolute velocities over 100 km s$^{-1}$) and Galactic coordinates. 
We derive interstellar extinction values for this sample using the position of the RGB in 2MASS fields near our maser sources. 
Additionally, these Mira variables should follow NIR period-magnitude relationships, and thereby their intrinsic NIR magnitudes can be determined from their periods assuming a distance of 8 kpc. We derive circumstellar extinction values for the sample using these intrinsic (i.e., period-derived) values and the aforementioned interstellar reddening values. The resulting circumstellar extinction values have a wide range, from practically no extinction up to 4 mag in the K$_s$ band. These results suggest a wide variety among CSE properties even for this small sample of sources all of which are likely in a similar environment (the Galactic bulge). 
\par Our study supports the idea that circumstellar extinction is a function of IR color, and shows that the correlations between K$_s$-band CSE extinction and J$-$A and K$_s-$A colors are most pronounced. As such, we suggest these colors can be used as a loose proxy for CSE extinction. 
Further, as samples curated from certain narrow J$-$A and/or K$_s-$A color ranges may be more homogeneous in terms of circumstellar properties, we also note that investigating PLRs in such samples constitutes important future work.  
\par Comparing circumstellar extinctions in the J, H, and K$_s$ bands shows they are strongly correlated, as expected. However, the relations between each pair of these extinction values do not follow standard interstellar reddening laws; instead they more closely match the circumstellar extinctions seen in Mira variables in the LMC in \cite{yuan17b}. We emphasize agreement between our values and O-rich LMC Miras in our empirical determination of A$_J$/A$_H$, A$_J$/A$_K$, and A$_H$/A$_K$, where we report values of 1.30, 1.75, and 1.35 respectively.
\par Finally, we present period-magnitude diagrams for bands covering 0.9-21 $\mu$m with interstellar extinction corrections applied. Positive correlations are not detectable for wavelengths $<$ 3 $\mu$m, but are clearly visible for longer wavelengths, and we present linear period-magnitude fits for WISE and MSX bands. It is clear that in WISE bands 2 and 3 the magnitude at a given period is generally brighter than predicted by LMC MIR relations given by \cite{iwanek21}. CSE effects can at least partially explain both the spread in MIR period-magnitude diagrams and the brighter magnitudes---as the CSE itself may contribute radiation at longer wavelengths. Likely SiO-maser samples, like the one explored here, have a wide variety of CSE conditions but tend toward very pronounced CSE effects.
OH-maser samples often dominate discussions of pronounced CSEs and high circumstellar extinction. However, even though SiO-maser bearing sources may have thinner envelopes in general than OH-maser stars, their envelopes can significantly affect measured magnitudes as shown here. 
There is also a great deal of overlap between SiO- and OH-bearing stars with many sources hosting both species of maser.
Moreover, I-band surveys such as OGLE are capable of detecting stars which host masers and have considerable circumstellar extinction, even at the distance of the GC.
We suggest that tactics such as creating separate PLRs for various K$_s-$A color bins should be considered to account for the role of the CSE when employing Mira PLRs for stars with any maser activity.

\begin{acknowledgements}
    We would like to thank the anonymous referee for helping us improve this manuscript.
      The research leading to these results has received funding from the European Research Council (ERC) Synergy "UniverScale" grant financed by the European Union's Horizon 2020 research and innovation programme under the grant agreement number 951549, from the National Science Center, Poland grants MAESTRO. We also acknowledge support from the DIR/WK/2018/09 grant of the Polish Ministry of Science and Higher Education.
      This work is supported by the National Aeronautics and Space Administration (NASA) under grant number 80NSSC22K0482 issued through the NNH21ZDA001N-ADAP Astrophysics Data Analysis Program (ADAP).
      This research made use of data products from the Midcourse Space Experiment. Processing of the data was funded by the Ballistic Missile Defense Organization with additional support from NASA Office of Space Science.
     This publication makes use of data products from the Two Micron All Sky Survey, which is a joint project of the University of Massachusetts and the Infrared Processing and Analysis Center/California Institute of Technology, funded by the National Aeronautics and Space Administration and the National Science Foundation.
     This publication makes use of data products from the Wide-field Infrared Survey Explorer, which is a joint project of the University of California, Los Angeles, and the Jet Propulsion Laboratory/California Institute of Technology, funded by the National Aeronautics and Space Administration.
\end{acknowledgements}

\bibliographystyle{aa} 
\bibliography{sample631.bib}

\end{document}